\documentclass[11pt,a4paper]{article}
\usepackage[margin=2.5cm]{geometry}
\usepackage{amsmath,amssymb}
\usepackage{graphicx}
\usepackage{booktabs}
\usepackage[T1]{fontenc}
\usepackage{lmodern}
\usepackage{xcolor}
\usepackage[colorlinks=true,linkcolor=black,citecolor=black,urlcolor=blue!55!black]{hyperref}
\usepackage{caption}
\newcommand{\bhat}{\hat{\boldsymbol b}}
\newcommand{\Khat}{\hat{\boldsymbol K}}
\newcommand{\nhat}{\hat{\boldsymbol n}}
\newcommand{\Bvec}{\boldsymbol B}
\newcommand{\rvec}{\boldsymbol r}
\newcommand{\svec}{\boldsymbol s}
\newcommand{\Bnorm}{\widetilde{B}_n}
\newcommand{\Kvec}{\boldsymbol K}
\newcommand{\taug}{\tau_g}
\newcommand{\taugB}{\tau_g^{(B)}}
\newcommand{\taugK}{\tau_g^{(K)}}

\title{Geodesic torsion of the magnetic field line and other predictors\\ of coil non-planarity in QI stellarators}
\author{%
  A. Pavone\textsuperscript{1}\thanks{Corresponding author: \texttt{andrea.pavone@ipp.mpg.de}} \and
  S. Kwak\textsuperscript{1} \and
  F. Warmer\textsuperscript{1}
}

\date{%
  \textsuperscript{1}Max-Planck-Institut f{\"u}r Plasmaphysik, Greifswald, Germany
}

\begin{document}
\maketitle

\begin{abstract}
Stellarator magnetic fields are often realized with non-planar modular coils with
their out-of-plane bending being one of the main engineering challenges. We ask whether the 
non planarity of the fialment can be predicted from the equilibrium alone. 
We look at a dataset of coils for stellarator equilibira and find that the geodesic torsion
of the magnetic field line is a strong and robust predictor in a wide range of coil feasibility requirements.
We compare the performance of the geodesic torsion against other equilibrium and surface geometry 
properties and find it to be a superior predictorv hardly replaceable by other quantities we considered in this dataset.
The ranking it induces also reproduces the ordering of four published
configurations --- precise QA, precise QI at $n_{fp}=2,3$, and W7-X. 
Our conclusions suggest that geodesic torsion of the magnetic field line
may be considered as a cheap and robust candidate for a stage-one non-planarity proxy.
\end{abstract}

\section{Introduction}
\label{sec:intro}

Stellarator equilibria require non-planar modular coils to produce their three-dimensional
magnetic fields. The non-planarity of those coils is a major driver of engineering cost and
one of the main risks in the design and construction of a stellarator power plant. With
several companies now pursuing power plants based on the stellarator
concept~\cite{stellaris,infinitytwo,giga}, it is timely to ask how the equilibrium configuration of a stellarator is related to the non-planarity of
the coils required to produce it.

This work follows a number of earlier studies of coil buildability for three-dimensional
equilibria. Landreman and Boozer~\cite{landremanboozer} show that the difficulty of
supporting a plasma shape with distant coils does not follow from any single feature of the
surface geometry: shapes of low curvature and low spectral width may be supported
inefficiently, while shapes with sharp edges may be supported efficiently. Kappel
\emph{et al.}~\cite{kappel2024,kappel2026} propose the minimum magnetic gradient scale
length as a boundary metric and relate it to the plasma--coil separation that a coil set
requires in order to reproduce the boundary field, the later work extending the result from
winding-surface current potentials to filamentary coils. Fu \emph{et
al.}~\cite{quadcoil,quadcoilproxy} show that a global winding-surface optimisation can serve
both as a measure of coil complexity and as a differentiable proxy for coil design within
single-stage optimisation. Hudson \emph{et al.}~\cite{hudson} compute derivatives of the
coil shapes with respect to the plasma boundary, which quantifies the sensitivity of the
coils to the boundary shape. Sengupta \emph{et al.}~\cite{sengupta} treat quasi-axisymmetric
and quasi-helically symmetric configurations and present a theory of the ridges that often
appear on their magnetic surfaces, and of how those ridges relate to coil complexity. Plunk
\emph{et al.}~\cite{plunk} report that a magnetic axis of low torsion can give lower
non-planarity, and present a figure-eight stellarator whose coils are close to planar.

The work of Rodr\'iguez and Sengupta~\cite{rs} is the most directly related to ours, as it
proposes a mechanism driving the non-planarity of the coils. Applying the virtual-casing
principle to the equilibrium at the boundary, they show that the non-planarity of the
current-density contours can be estimated from the equilibrium alone. We follow that work by
providing a link between the geodesic torsion of the magnetic field lines and the
non-planarity of filamentary coils at standoff, and by testing it on a dataset of coil sets.
The same quantity appears in the plasma literature under the name normal
torsion~\cite{skovoroda,hegna}.
What this study adds is the scale of that test: the relation is examined across a dataset of
QI-like equilibria and their filamentary coils, with the statistical validation carried out
across field-period classes and constraint sets.

Section~\ref{sec:methods} describes the dataset and the quantities used in the analysis,
Sec.~\ref{sec:results} presents the trends between the candidate predictors and the coil
non-planarity, and Sec.~\ref{sec:conclusions} discusses the implications and the work that
follows from them.

\section{Methods}
\label{sec:methods}

\subsection{The dataset}
\label{sec:dataset}

The dataset is described in detail in~\cite{paper1}; here we summarise only what the present
analysis requires. Stage-one QI boundaries are taken from the ConStellaration
database~\cite{constellaration}. We retain only vacuum equilibria with a stabilising vacuum
well, that is a positive vacuum well in the convention of the database, and with between two
and five field periods. This leaves $3836$ configurations, which we refer to as the
\emph{dataset}. Table~\ref{tab:nfp-population} gives the number of configurations for each
field-period count.

These boundaries belong to two provenance families, as described in~\cite{constellaration}.
The first consists of boundaries carrying poloidal modes $m\le2$, generated by near-axis or
quasi-poloidal heuristics and intended as initial guesses for the fixed-boundary
optimisation. The second consists of boundaries with modes up to $m=4$, obtained from
fixed-boundary optimisation with DESC or VMEC++. The distinction matters here because the
two families support different trends. We therefore restrict the study to the second family, which is
the more relevant of the two for the design of a stellarator power plant; the first is not
considered further. Further details of the stage-one fixed-boundary optimisation behind the second
family, including its targets and constraints, are given in~\cite{constellaration}.

\begin{table}[htbp]
  \centering
  \caption{Number of configurations in the dataset for each field-period count, split by
  provenance family: boundaries generated as initial guesses by the near-axis or
  quasi-poloidal heuristics, and boundaries obtained from fixed-boundary optimisation with
  DESC or VMEC++. The dataset is the subset of ConStellaration~\cite{constellaration} with a
  positive vacuum well. The analysis that follows uses the optimised family.}
  \label{tab:nfp-population}
  \setlength{\tabcolsep}{10pt}
  \renewcommand{\arraystretch}{1.2}
  \begin{tabular}{c c c c}
    \toprule
    $n_{fp}$ & initial-guess family & optimised family & total \\
    \midrule
    2   & 398 &  862 & 1260 \\
    3   & 302 & 1604 & 1906 \\
    4   &  25 &  519 &  544 \\
    5   &   1 &  125 &  126 \\
    \midrule
    all & 726 & 3110 & 3836 \\
    \bottomrule
  \end{tabular}
\end{table}

\begin{figure}[htbp]
  \centering
  \includegraphics[width=\linewidth]{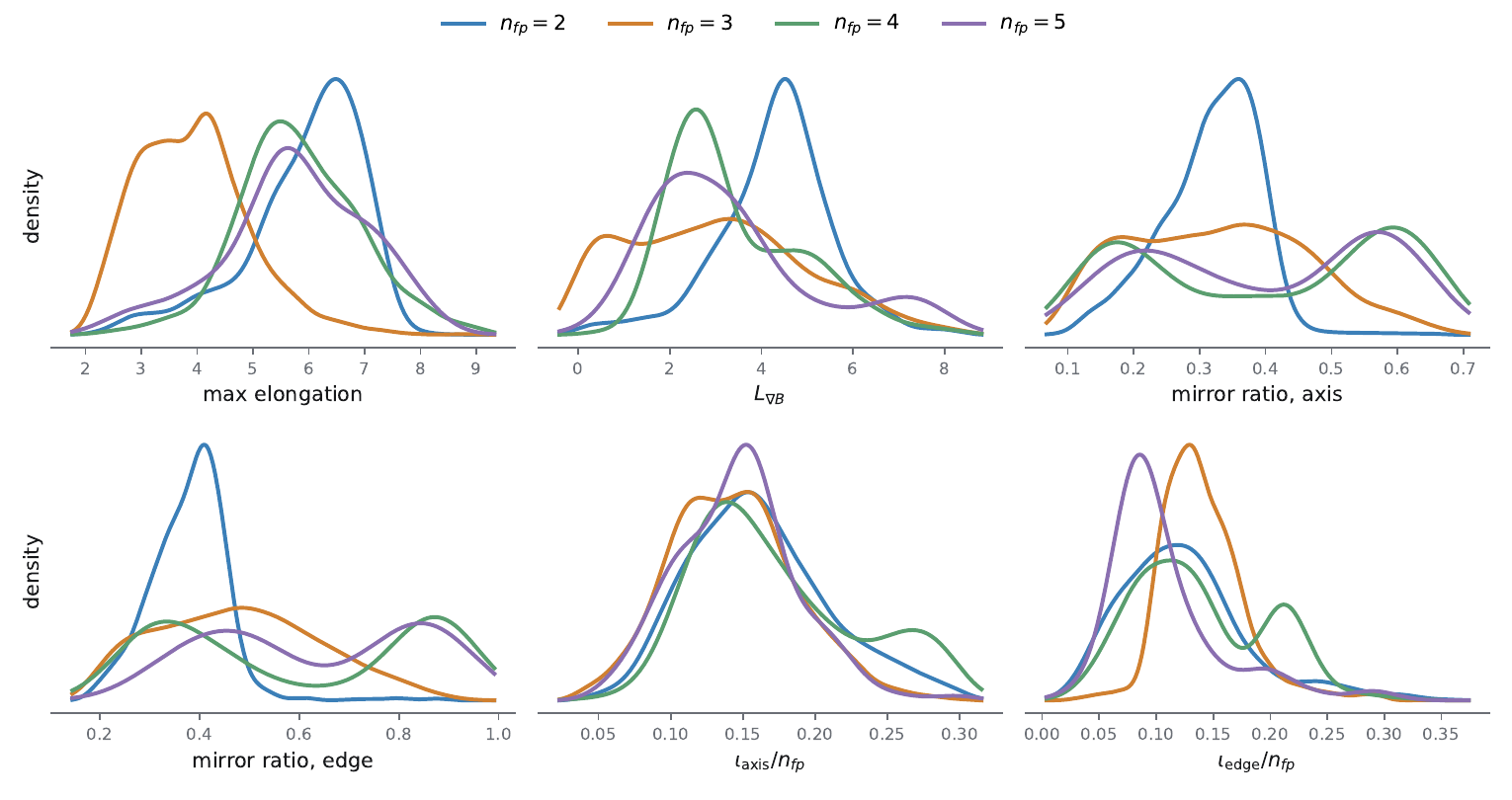}
  \caption{Distribution of the stage-one physics quantities over the dataset, by field-period
  count. Values are taken from the database's own metrics~\cite{constellaration}.}
  \label{fig:dataset-dists}
\end{figure}

Figure~\ref{fig:dataset-dists} shows how the stage-one physics quantities are distributed
over this population, separated by field-period count. The ranges are wide and differ
systematically between field-period counts: the $n_{fp}=2$ boundaries carry the largest
elongations and the smallest edge mirror ratios, while $n_{fp}=4$ and $5$ reach edge mirror
ratios above $0.8$. This is the covariation referred to in Sec.~\ref{sec:predictors}, and the
reason the correlations of Sec.~\ref{sec:results} are computed within a field-period class
rather than across the population as a whole.

For each boundary, filament coils are optimised with \textsc{simsopt}~\cite{simsopt}. The
details are given in~\cite{paper1}; we reproduce here the quantities the present analysis
refers to. The optimisation uses an augmented-Lagrangian method~\cite{gilprl,gilpre} which
drives five penalty terms to zero. The field is reconstructed through the normalised squared
flux over the boundary $S$,
\begin{equation}
  J_B \;=\; \frac{1}{2}\,
  \frac{\int_S (\Bvec\cdot\nhat)^2\,\mathrm{d}s}{\int_S |\Bvec|^2\,\mathrm{d}s},
  \label{eq:jb}
\end{equation}
the normalisation preventing the optimiser from lowering the objective by weakening the
field. The two clearances are imposed as the one-sided integrals of the \textsc{simsopt}
convention,
\begin{align}
  J_{cs}^{(p)} &= \sum_i \int_{C_i}\!\int_S
     \max\!\left(0,\; d_{cs} - \|\rvec_i - \svec\|\right)^p \mathrm{d}s\,\mathrm{d}l_i ,
  \label{eq:jcs}\\
  J_{cc}^{(p)} &= \sum_i \sum_{j<i} \int_{C_i}\!\int_{C_j}
     \max\!\left(0,\; d_{cc} - \|\rvec_i - \rvec_j\|\right)^p \mathrm{d}l_j\,\mathrm{d}l_i ,
  \label{eq:jcc}
\end{align}
over the coil curves $C_i$, and the curvature by the $L_p$ penalty
\begin{equation}
  J_\kappa^{(p)} \;=\; \sum_i \frac{1}{p}\int_{C_i}
     \max\!\left(0,\; \kappa_i - k_{\max}\right)^{p} \mathrm{d}l_i ,
  \label{eq:jk}
\end{equation}
and the total length by
\begin{equation}
  J_L \;=\; \max\!\left(0,\; \textstyle\sum_i L_i - L_t\right)^2 ,
  \qquad
  L_t \;=\; (1+\ell)\,2\pi\,(a + d_{cs})\,n_c ,
  \label{eq:jl}
\end{equation}
where $\ell$ is the length target ratio, $a$ the minor radius and $n_c$ the number of unique
coils per half period. The augmented Lagrangian uses $p=2$ throughout.
Every one of \eqref{eq:jcs}--\eqref{eq:jl} is one-sided, so driving it
to zero enforces the corresponding inequality --- $\|\rvec_i-\svec\|\ge d_{cs}$ and
$\|\rvec_i-\rvec_j\|\ge d_{cc}$ everywhere along the filaments, $\kappa_i\le k_{\max}$, and
$\sum_i L_i \le L_t$ --- rather than an equality. Treating them this way lets the optimiser
find the Lagrange multipliers on its own, without the manual weight tuning that would
otherwise make a dataset of this size impractical to produce.

Two consequences matter for what follows. The minimum coil--surface and coil--coil distances
and the maximum curvature quoted later are diagnostics measured on the converged coil set;
the constrained quantities are the integrals \eqref{eq:jcs}--\eqref{eq:jk} themselves, and a
small residual value of one of them corresponds to a violation confined to a short stretch of
filament rather than to a uniform shortfall. The quality of the reconstruction is reported
throughout as the normal field error
\begin{equation}
  \Bnorm \;=\; \frac{\langle\,|\Bvec\cdot\nhat|\,\rangle}{\langle\,|\Bvec|\,\rangle},
  \label{eq:fielderr}
\end{equation}
with $\langle\cdot\rangle$ the mean over the boundary quadrature points, which is a more
directly interpretable measure of the residual normal field than \eqref{eq:jb}. Across the
dataset we use the pointwise-normalised variant $\langle|\Bvec\cdot\nhat|/|\Bvec|\rangle$, which
rank-correlates at $0.990$ with \eqref{eq:fielderr} and is $0.96$ times as large, so the same
threshold serves for both.

The two minimum distances and the maximum curvature are varied over seven constraint sets,
and each set is run at several length targets and several numbers of coils per half period,
as listed in Table~\ref{tab:constraint-grid}. The sets are chosen so that each of the three
bounds moves on its own at least once, rather than only in step with the others: $d_{cc}$
varies at fixed $d_{cs}$, $d_{cs}$ at fixed $d_{cc}$, and $k_{\max}$ at both distances held
fixed. This gives fifty-four campaigns. Within a
campaign every boundary is optimised under the same constraints, which allows a fair
comparison across boundaries.

\begin{table}[htbp]
  \centering
  \caption{The constraint sets over which the dataset was produced. The two distances are
  given as multiples of the minor radius and the curvature bound as $k_{\max}a$; the length
  target ratio $\ell$ is defined by Eq.~\eqref{eq:jl}. Each line is one campaign per
  listed $n_c$, fifty-four in all. Lines marked $\dagger$ were run on the $n_{fp}=4,5$
  boundaries only; the rest cover $n_{fp}=2$--$5$. The two sets at $d_{cc}/a=0.17$ with
  $d_{cs}/a=0.83$ and at $d_{cc}/a=0.25$ with $d_{cs}/a=1.00$ exist so that the two
  distances are not varied only in step with one another, and the rows at
  $k_{\max}a=2.5$ and $4$ hold both distances at $0.83$ and $0.25$ so that the
  curvature bound likewise moves on its own.}
  \label{tab:constraint-grid}
  \setlength{\tabcolsep}{9pt}
  \renewcommand{\arraystretch}{1.2}
  \begin{tabular}{c c c c c}
    \toprule
    $d_{cs}/a$ & $d_{cc}/a$ & $k_{\max}a$ & length target ratio & $n_c$ \\
    \midrule
    0.83 & 0.17 & 3 & 0.6 & 5, 6 \\
    0.83 & 0.17 & 3 & 0.7 & 5, 6 \\
    \addlinespace
    0.83 & 0.25 & 2.5 & 0.6 & 5, 6 \\
    0.83 & 0.25 & 2.5 & 0.9 & 5, 6 \\
    \addlinespace
    0.83 & 0.25 & 3 & 0.5 & 5, 6, 7, 8$^\dagger$ \\
    0.83 & 0.25 & 3 & 0.6 & 5, 6, 7, 8$^\dagger$ \\
    0.83 & 0.25 & 3 & 0.7 & 5, 6, 7, 8 \\
    0.83 & 0.25 & 3 & 0.8 & 5, 6, 7$^\dagger$ \\
    0.83 & 0.25 & 3 & 0.9 & 5, 6, 7, 8 \\
    \addlinespace
    0.83 & 0.25 & 4 & 0.6 & 5, 6 \\
    0.83 & 0.25 & 4 & 0.9 & 5, 6 \\
    \addlinespace
    1.00 & 0.17 & 3 & 0.6 & 4 \\
    1.00 & 0.17 & 3 & 0.7 & 4, 5, 6, 7, 8 \\
    1.00 & 0.17 & 3 & 0.8 & 4 \\
    1.00 & 0.17 & 3 & 0.9 & 4 \\
    \addlinespace
    1.00 & 0.25 & 3 & 0.6 & 5, 6 \\
    1.00 & 0.25 & 3 & 0.7 & 5, 6 \\
    \addlinespace
    1.17 & 0.33 & 2 & 0.6 & 5, 6, 7 \\
    1.17 & 0.33 & 2 & 0.7 & 5, 6, 7, 8 \\
    1.17 & 0.33 & 2 & 0.8 & 5, 6, 7, 8 \\
    \bottomrule
  \end{tabular}
\end{table}

To exclude solutions in which the optimiser has not converged we apply a feasibility filter.
It is built from the same three penalties evaluated at $p=1$ and normalised by their own
thresholds,
\begin{equation}
  \frac{J_{cs}^{(1)}}{d_{cs}} ,\qquad
  \frac{J_{cc}^{(1)}}{d_{cc}} ,\qquad
  \frac{J_\kappa^{(1)}}{k_{\max}} \;<\; 10^{-2} ,
  \label{eq:filter}
\end{equation}
required together with $\Bnorm < 5\times10^{-3}$ and $\sum_i L_i \le 1.02\,L_t$. The exponent is
lowered because at $p=2$ a shallow violation is suppressed quadratically, which makes
\eqref{eq:jcs}--\eqref{eq:jk} poor detectors of one; at $p=1$ the measure is linear in the
depth of the violation. The filter is nonetheless the more tolerant criterion of the two,
since the optimiser was asked to drive the penalties to zero whereas \eqref{eq:filter} admits
up to $10^{-2}$. Being integrals, these measures do not necessarily exclude coil sets whose
clearance falls short at a small number of locations along the filaments.

\subsection{Coil non-planarity}
\label{sec:np}

Coils are modeled as 1d curve in 3D cartesian coordinates. The out-of-plane tilt of 
a curve can be measured with the torsion of the Frenet frame which has the drawback
of not being defined where curvature is vanishing. 
Therefore we measure non planarity globally as the overall deviation from the best fit
plane to the curve.
For each coil we take its quadrature points $P\in\mathbb{R}^{N\times3}$, subtract the
centroid, and compute the singular values $\sigma_1\ge\sigma_2\ge\sigma_3$ of the centred
matrix. The non-planarity score is
\begin{equation}
  \mathrm{NP} \;=\; \frac{\sigma_3}{\sigma_1+\sigma_2+\sigma_3},
  \label{eq:np}
\end{equation}
since $\sigma_3$ measures the spread orthogonal to the best-fit plane.
\begin{figure}[htbp]
  \centering
  \includegraphics[width=\linewidth]{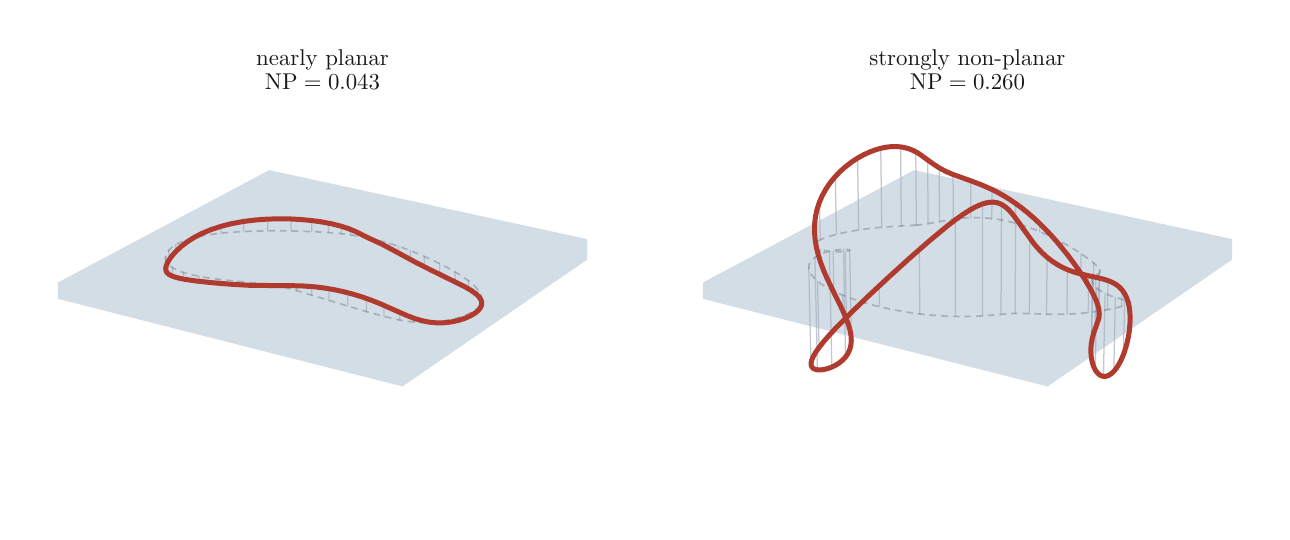}
  \caption{Two coils from the dataset with their best-fit planes: one nearly planar
  (left) and one strongly non-planar (right), the latter at the $99$th percentile of the
  non-planarity found over the feasible population of its campaign.}
  \label{fig:np-lowhigh}
\end{figure}
The score is zero for a planar coil and it is dimensionless therefore adequate for a 
cross-configuration comparison. 
The figure of merit for a coil set is the mean of~\eqref{eq:np} over its coils.
Figure~\ref{fig:np-lowhigh} shows two coil filament, one mostly planar and the other one 
bending strongly out-of-plane, together with the best fit plane and the
value of non-planarity. 

\subsection{The geodesic torsion on the flux surface}
\label{sec:theory}

Following~\cite{rs}, the current density on the last closed
flux surface of a vacuum equilibrium is
\begin{equation}
  \Kvec \;=\; -\frac{1}{\mu_0}\,\nhat \times \Bvec
  \;=\; \frac{1}{\mu_0}\,\nhat \times \nabla\Phi ,
  \label{eq:rs}
\end{equation}
with $\Phi$ the current potential. The second form matters: $\Kvec$ is tangent to the
contours of $\Phi$, and those contours are idealised coils whose out-of-plane excursion
drives the out-of-plane excursion of coil filaments at standoff.

For any curve on a surface the Frenet torsion splits into two terms~\cite{docarmo},
\begin{equation}
  \tau^{(K)} \;=\; \taugK \;+\; \frac{\mathrm{d}\theta}{\mathrm{d}s} ,
  \label{eq:split}
\end{equation}
where $\taugK$ is the geodesic torsion of the $\Kvec$-line, $\theta$ is the angle between
the curve normal $\boldsymbol N$ and the surface normal $\nhat$, and $s$ is arclength along
the $\Kvec$-line. The Frenet frame, and hence $\theta$, is undefined wherever the curvature
of the curve vanishes, so $\mathrm{d}\theta/\mathrm{d}s$ is singular at those points and
cannot be evaluated as a field over the whole surface, whereas $\taugK$ can. We ask instead
whether the first term alone is a sufficient proxy for the non-planarity of the
$\Kvec$-line.

The geodesic torsion of a curve on a surface depends only on its direction~\cite{docarmo},
\begin{equation}
  \taug \;=\; \tfrac{1}{2}\,(k_1 - k_2)\,\sin 2\alpha ,
  \label{eq:bonnet}
\end{equation}
with $k_1$ and $k_2$ the principal curvatures and $\alpha$ the angle between the tangent and
the $k_1$ direction. In practice $\taug$ is evaluated in the equivalent invariant form, which
needs neither the principal directions nor $\alpha$; Appendix~\ref{app:taug-numerics} gives it
and the corresponding calls. By \eqref{eq:rs} the vector $\Kvec$ is perpendicular to $\Bvec$ and
tangent to the surface, so replacing $\alpha$ by $\alpha+\pi/2$ gives $\taugK = -\taugB$ and
\begin{equation}
  \tau^{(K)} \;=\; -\,\taugB \;+\; \frac{\mathrm{d}\theta}{\mathrm{d}s} ,
  \label{eq:tauK}
\end{equation}
so the torsion of the current contours follows from the flux surface and the field on it
alone.

The surface quantity $\taugK$ also fixes, to first order, how the current contours continue
off the surface, which is the extrapolation Ref.~\cite{rs} notes their construction
requires before it can be compared with coils at finite standoff. Differentiating
\eqref{eq:rs} along the normal,
\begin{equation}
  \partial_n\Kvec \;=\; -\frac{1}{\mu_0}\Big[(\partial_n\nhat)\times\Bvec + \nhat\times\partial_n\Bvec\Big] .
  \label{eq:dnK}
\end{equation}
Since $|\nhat|=1$, $\partial_n\nhat$ is tangent to the surface, so the first term is directed
along $\nhat$. Writing $\partial_n\Bvec$ in the frame $(\bhat,\Khat,\nhat)$, the cross
product in the second term discards the normal component and turns the other two by a right
angle,
\begin{equation}
  \partial_n\Kvec \;=\; \frac{1}{\mu_0}\Big[(\bhat\cdot\partial_n\Bvec)\,\Khat
  \;-\; (\Khat\cdot\partial_n\Bvec)\,\bhat\Big]
  \;-\;\frac{1}{\mu_0}\,(\partial_n\nhat)\times\Bvec .
  \label{eq:dnK-frame}
\end{equation}
The three components have different meanings. The component along $\Khat$ changes the
magnitude of $\Kvec$ and not its direction. The component along $\bhat$ rotates $\Kvec$ about
the normal and therefore alters the shape of the contour within the surface; in the vacuum
region it is the geodesic torsion. There $\nabla\times\Bvec = 0$, i.e.\
$\partial_iB_j = \partial_jB_i$, so the two directions of the mixed derivative may be
exchanged, $\Khat\cdot(\nhat\cdot\nabla)\Bvec = \nhat\cdot(\Khat\cdot\nabla)\Bvec$; the
right-hand side is a derivative along the current contour, and since $\nhat\cdot\bhat=0$
only the change of direction survives, for which the Darboux frame gives
$\nhat\cdot(\Khat\cdot\nabla)\bhat = \taugK$. Hence
\begin{equation}
  \partial_n\Kvec \;=\; \frac{\partial_n|\Bvec|}{\mu_0}\,\Khat \;-\; |\Kvec|\,\taugK\,\bhat
  \;-\;\frac{1}{\mu_0}\,(\partial_n\nhat)\times\Bvec .
  \label{eq:continuation}
\end{equation}
The last component, along $\nhat$,
depends on how the family of surfaces away from the boundary is constructed. For parallel
surfaces, obtained by carrying each point a fixed distance along its own normal,
$\partial_n\nhat=0$ and the term vanishes. For the family of flux surfaces, with
$\nhat=\nabla\psi/|\nabla\psi|$, it is in general not negligible.
It describes on which surface the continued contour lies.
It also tilts the tangent plane that the current vectors occupy as one steps off the
surface, and since that tilt need not be uniform along a contour it may contribute to
the out-of-plane excursion of the filament curves at finite standoff; we do not pursue
it here.
The rotation of the contour within the tangent plane is set by the $\bhat$ component
alone, so the geodesic
torsion is the leading term in the normal expansion of the current contours away from the
flux surface, and $\mathcal{T}$ of \eqref{eq:T} is the area-weighted magnitude of that term.

By \eqref{eq:bonnet}, $\taugB$ vanishes where the field line runs along a line of curvature,
as on a torus of revolution carrying a purely toroidal field and no rotational transform,
and at umbilic points, where $k_1 = k_2$. It is large where $k_1 - k_2$ is large, or where
the field line is strongly misaligned with the lines of curvature.

\begin{figure}[htbp]
  \centering
  \includegraphics[width=0.48\linewidth]{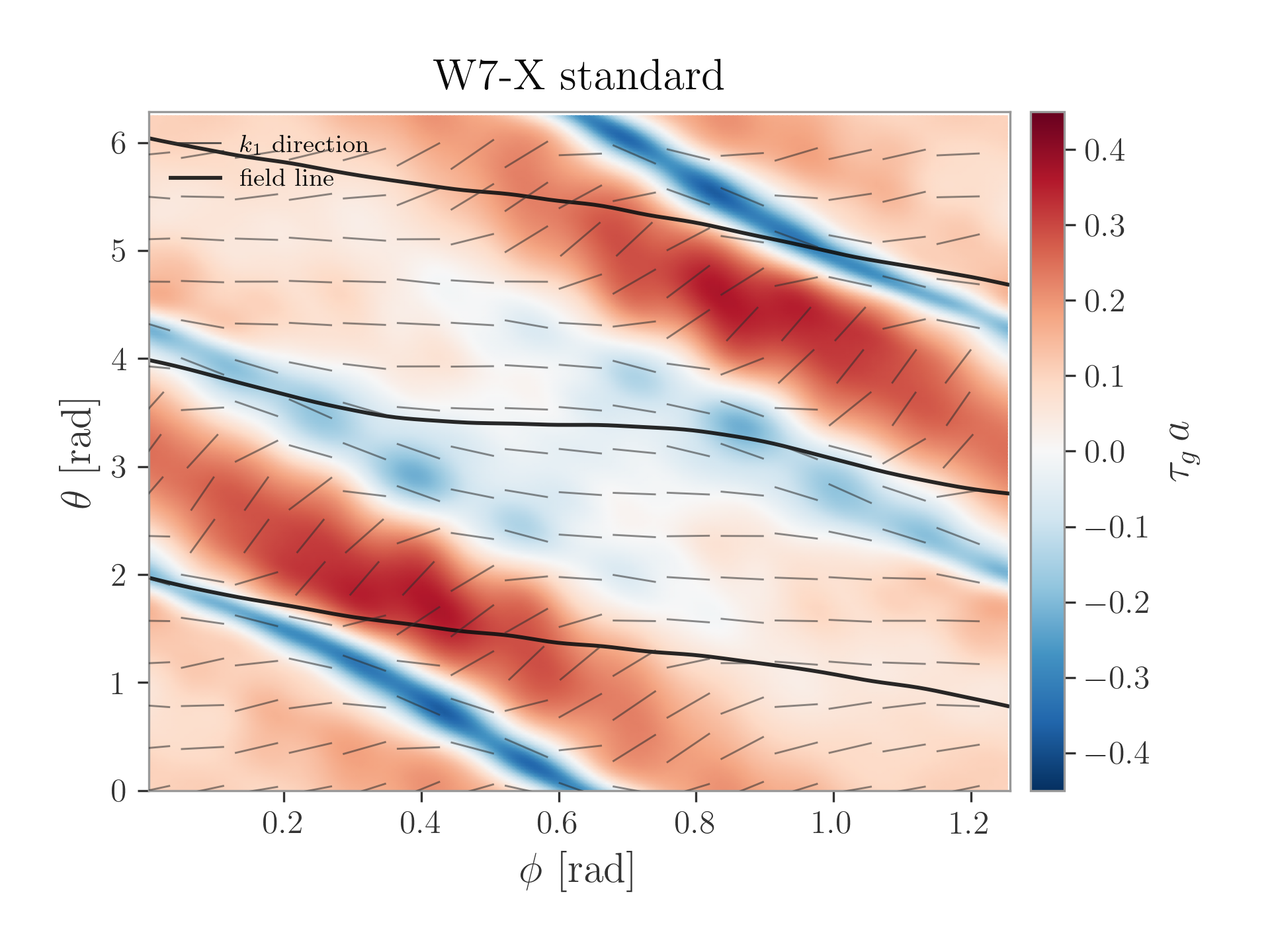}
  \includegraphics[width=0.48\linewidth]{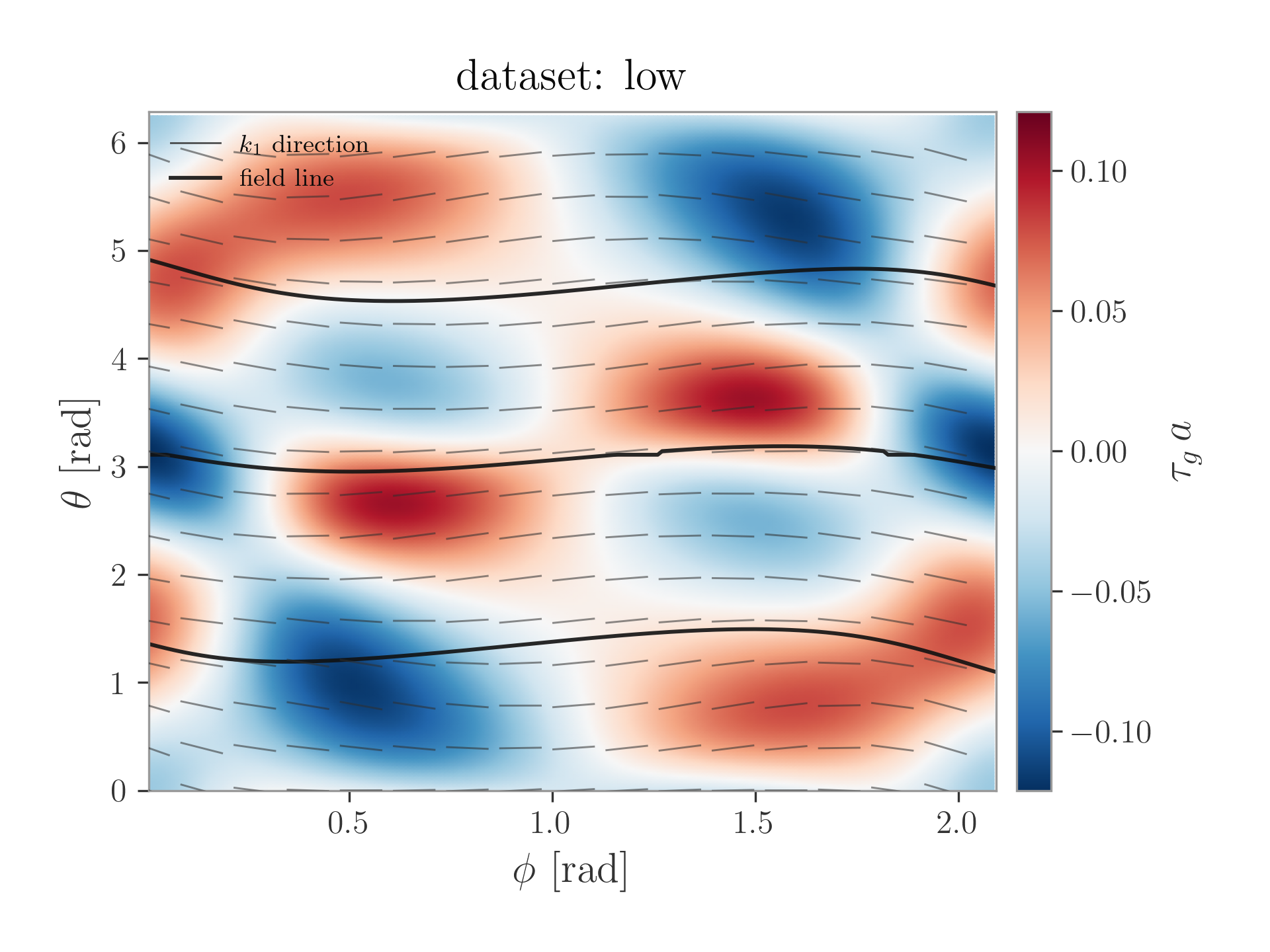}\\[4pt]
  \includegraphics[width=0.48\linewidth]{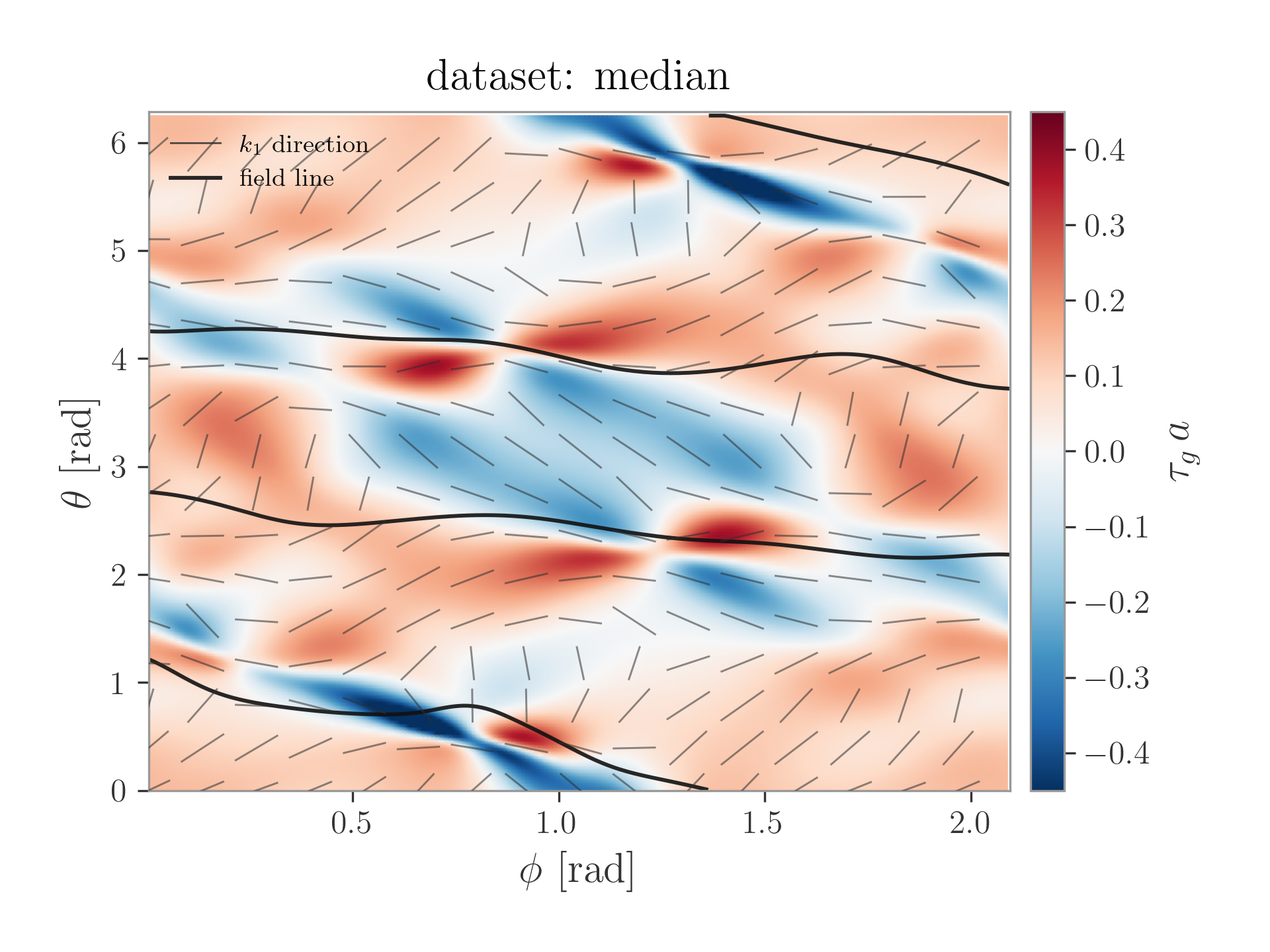}
  \includegraphics[width=0.48\linewidth]{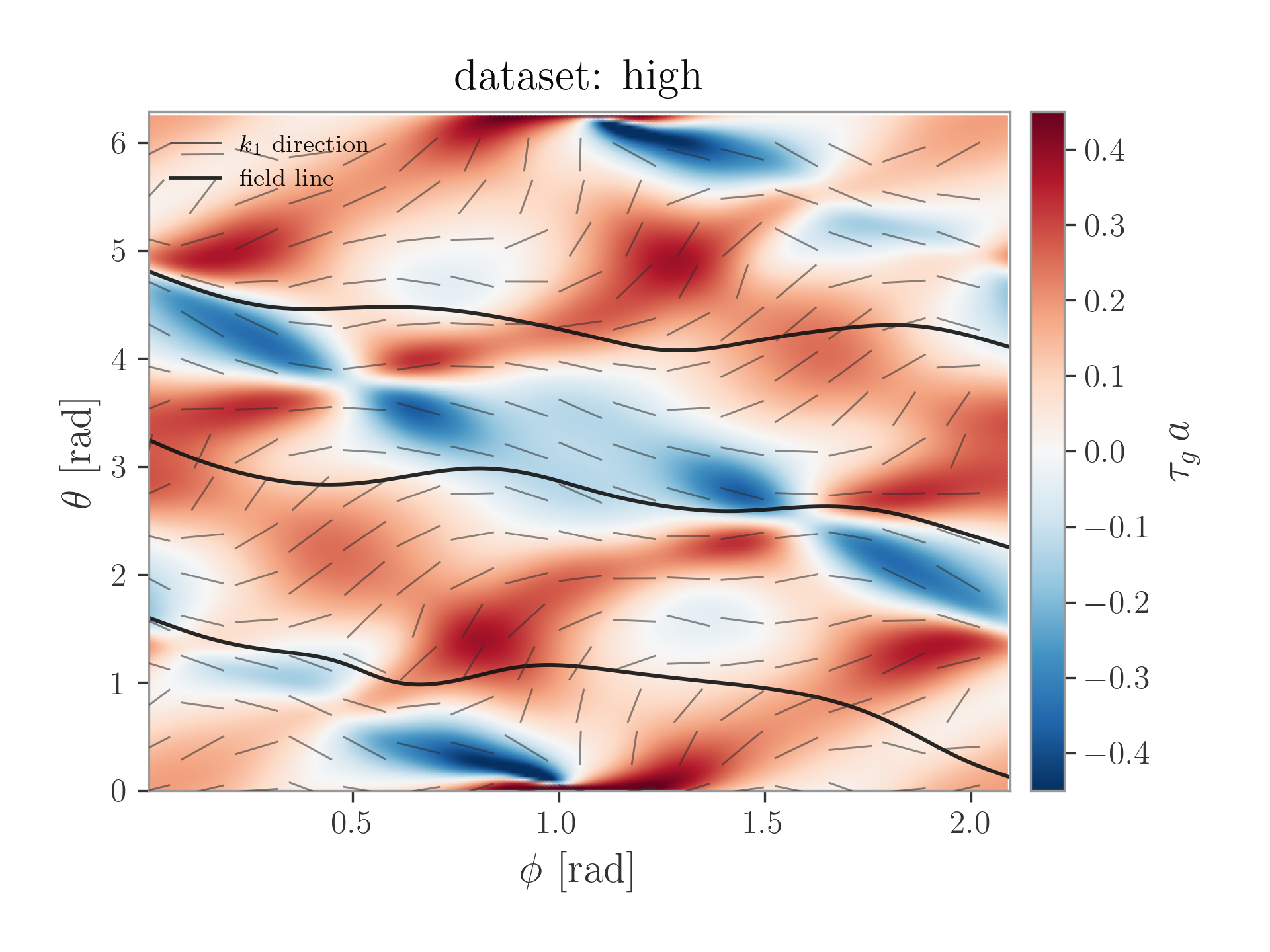}
  \caption{Field-line geodesic torsion $\taugB a$ on the plasma boundary, over one full field
  period $\phi \in [0, 2\pi/n_{fp})$. \textbf{Top left:} W7-X standard vacuum configuration 
  ($n_{fp}=5$, $\mathcal{T}=0.132$). \textbf{Top right, bottom left, bottom right:} 
  three $n_{fp}=3$ dataset configurations spanning the $\mathcal{T}$ distribution, at its
  lower end, near its median and in its upper range ($\mathcal{T} = 0.039$, $0.106$,
  $0.135$). The low-torsion
  panel (top right) has its own scale. The high-torsion cases show sharp 
  sign-reversing bands where the field line crosses the principal directions.
  Overlaid are three field lines (solid) and the $k_1$ principal direction (thoroidal) 
  as unoriented ticks. 
  The pale bands are where the field line is close to a line of curvature, 
  which by Eq.~\eqref{eq:bonnet} is one of the two ways $\taug$ can vanish.}
  \label{fig:maps}
\end{figure}

Figure~\ref{fig:maps} shows $\taugB a$ on the boundary of four configurations: W7-X in its
standard vacuum configuration, and three $n_{fp}=3$ members of the dataset spanning the
distribution of the scalar $\mathcal{T}$ defined below in \eqref{eq:T} --- one near its
lower end, one near its median, and one in its upper range. Each map covers one full field period. The quantity
is signed and $\taug$ vanishes wherever the field line runs along a line of
curvature.

\begin{figure}[htbp]
  \centering
  \includegraphics[width=\linewidth]{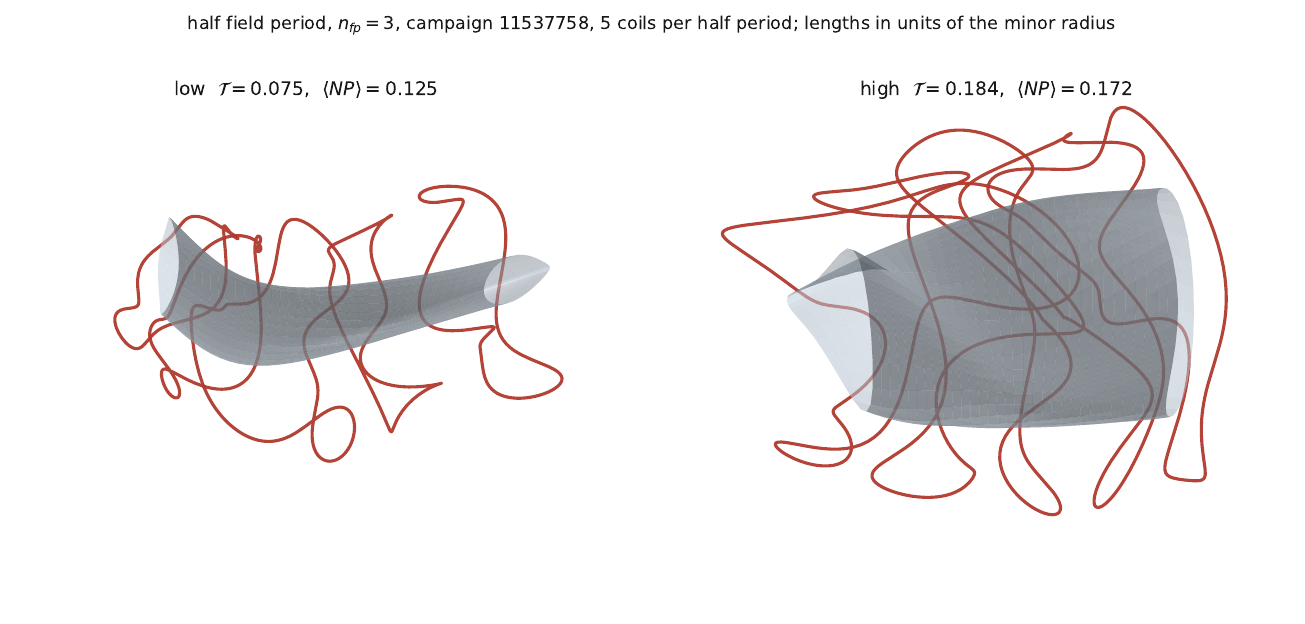}
  \caption{Plasma boundary and filament coils over one half field period, for a low- and a
  high-$\mathcal{T}$ configuration at $n_{fp}=3$ with $n_c=5$
  coils per half period. Both lie in the feasible population.}
  \label{fig:coils3d}
\end{figure}

Fig.~\ref{fig:coils3d} shows two configurations of differing $\mathcal{T}$, to illustrate how
the coil shapes differ under the same optimisation settings. Both are $n_{fp}=3$ configurations
from the same campaign, so thecoil optimiser was given the identical constraint set in both 
cases and the only thing that differs is the plasma shape. The low-$\mathcal{T}$
boundary is close to a surface of revolution over the half period and its coils are close to
planar; the high-$\mathcal{T}$ boundary twists through the period and every coil follows it
out of plane.

$\taugB$ is defined pointwise on the flux surface. To obtain one scalar per stage-one
boundary we take its area-weighted mean magnitude, made dimensionless with the minor radius,
\begin{equation}
  \mathcal{T} \;=\; a\,
  \frac{\displaystyle\int_S |\taugB|\,\mathrm{d}S}{\displaystyle\int_S \mathrm{d}S} .
  \label{eq:T}
\end{equation}
The normalisation by $a$ is one choice among several; we adopt it because it makes
$\mathcal{T}$ comparable across surfaces of different size.

\subsection{Surface geometry and other magnetic predictors}
\label{sec:predictors}

$\mathcal{T}$ is one of a family of quantities that can be read off a stage-one equilibrium,
and in these configurations they are not independent of one another. The boundary shape fixes
the surface curvatures, the field on it and the shape of the magnetic axis together, so a
configuration with large $\mathcal{T}$ will as a rule also have large $|H|$, large
$|\kappa_n|$ and a strongly curved axis. A correlation between the coil non-planarity and any
one of them therefore cannot on its own say whether that quantity carries information of its
own or is standing in for a neighbour.

We treat this by computing two sets of coefficients. The first is the non-planarity against
every candidate, which ranks them (Sec.~\ref{sec:other-predictors}). The second is
$\mathcal{T}$ against every candidate (Appendix~\ref{app:taug-pairs}), which bounds how much
of a competing ranking is shared information rather than an independent trend: a quantity
that follows the non-planarity about as closely as $\mathcal{T}$ does, and is itself strongly
correlated with $\mathcal{T}$, is not a rival explanation but the same one written in
different variables.

Table~\ref{tab:predictors} lists those included in this study. They fall into four groups.

The first two are fields on the boundary: the differential geometry of the surface itself,
and the geometry of the field line in its Darboux frame, of which $\taugB$ is one member.
Both are reduced to one number per configuration by the area-weighted mean of their
magnitude and made dimensionless with the minor radius, exactly as in \eqref{eq:T}; where
the trends were computed signed quantities, 
the signed reduction is reported alongside the absolute one. The
third group describes the magnetic axis and is reduced by a plain mean along the curve. The
fourth are scalars supplied with the equilibrium configurations from Constellaration, already dimensionless and needing no
reduction.

The normal and geodesic curvatures $\kappa_n$ and
$\kappa_g$ are the two components of the field-line curvature vector, normal and tangent to
the surface; together with $\taug$ they are the full Darboux description of how a field line
sits on the boundary.

Three of these quantities are not independent. For any line on a surface~\cite{skovoroda},
\begin{equation}
  \kappa_n^2 \;+\; \taug^2 \;-\; 2H\kappa_n \;+\; K \;=\; 0 ,
  \label{eq:mohr}
\end{equation}

\begin{table}[htbp]
  \centering
  \caption{Quantities compared with $\mathcal{T}$. Fields on the boundary and quantities
  along the axis are reduced to one number per configuration as described in the text and
  made dimensionless with the minor radius $a$; the scalars in the last group are supplied
  with the equilibrium~\cite{constellaration} and are already dimensionless. $\alpha$ is the
  angle between the field line and the $k_1$ principal direction, as in \eqref{eq:bonnet}.}
  \label{tab:predictors}
  \setlength{\tabcolsep}{7pt}
  \renewcommand{\arraystretch}{1.25}
  \begin{tabular}{l l l}
    \toprule
    symbol & name & definition \\
    \midrule
    \multicolumn{3}{l}{\textit{surface differential geometry}} \\
    $k_1,\,k_2$      & principal curvatures  & eigenvalues of the shape operator \\
    $H\,a$           & mean curvature        & $H = \tfrac{1}{2}(k_1 + k_2)$ \\
    $|K|\,a^2$       & Gaussian curvature    & $K = k_1 k_2$ \\
    \addlinespace
    \multicolumn{3}{l}{\textit{field-line geometry on the boundary (Darboux frame)}} \\
    $\kappa_n\,a$    & normal curvature      & $\kappa_n = k_1\cos^2\alpha + k_2\sin^2\alpha$ \\
    $\kappa_g\,a$    & geodesic curvature    & in-surface part of the curvature vector \\
    $\mathcal{T}$    & geodesic torsion      & $\taug = \tfrac{1}{2}(k_1-k_2)\sin 2\alpha$,
                                               reduced by \eqref{eq:T} \\
    \addlinespace
    \multicolumn{3}{l}{\textit{magnetic axis}} \\
    $\kappa_{\rm axis}\,a$ & axis curvature  & Frenet curvature of the magnetic axis \\
    $\tau_{\rm axis}\,a$   & axis torsion    & Frenet torsion of the magnetic axis \\
    \addlinespace
    \multicolumn{3}{l}{\textit{scalars supplied with the equilibrium}} \\
    $q_i$                        & quasi-isodynamicity residual & departure from omnigeneity \\
    $L_{\nabla B}$               & magnetic gradient scale length
                                 & $\sqrt{2}\,|\Bvec|/\|\nabla\Bvec\|_F$, minimum on the
                                   boundary~\cite{kappel2024} \\
    vacuum well                  & magnetic well depth          & stabilising for positive values \\
    $\iota_{\rm axis}/n_{fp}$    & rotational transform on axis & per field period \\
    $\iota_{\rm edge}/n_{fp}$    & rotational transform at edge & per field period \\
    mirror$_{\rm axis}$          & mirror ratio on axis         & $(B_{\max}-B_{\min})/(B_{\max}+B_{\min})$ \\
    mirror$_{\rm edge}$          & mirror ratio at edge         & as above, on the boundary \\
    max elongation               & cross-section elongation     & maximum over the field period \\
    \bottomrule
  \end{tabular}
\end{table}

\section{Results}
\label{sec:results}

\subsection{Correlation of $\mathcal{T}$ with coil non-planarity across the dataset}
\label{sec:dataset-results}

We computed $\mathcal{T}$ and the mean coil non-planarity~\eqref{eq:np} for every coil set in
the feasible population of all campaigns. All coefficients are computed within
a single $(\text{campaign},\,n_{fp})$ cell, so they measure variation inside
a field-period class at fixed constraint set. Wherever cells are afterwards combined, both
quantities are first \emph{demeaned}: the mean of a cell is subtracted from each of its
members, so that every cell is centred on zero and only the variation within it is retained.
Demeaned quantities therefore take both signs, and their zero marks the cell average rather
than any absolute value.

Table~\ref{tab:dataset} reports, for each $n_{fp}$, the distribution of Spearman and
Pearson coefficients across all campaign cells reaching sample size $n\ge 50$. A cell is one
$(\text{campaign},\,n_{fp})$ pair: the fifty-four campaigns of Table~\ref{tab:constraint-grid}
contain boundaries at several field-period counts each, giving $183$ such pairs, of which
$92$ reach $n\ge50$. 
We summarise the cells rather than list them per campaign deliberately: 
the claim is not about any one constraint set but about the stability of the relation across 
all of them. The median and interquartile  range describe the typical cell; 
the full range is given in parentheses because the weakest cell is itself part of the claim. 
Note that the extremes of a sample widen as more cells are added, whereas the quartiles do not, 
so the interquartile range is the quantity to compare between rows.

\begin{table}[htbp]
  \centering
  \caption{Correlation between $\mathcal{T}$ and mean coil non-planarity, feasible
  population, demeaned within $(\text{campaign},n_{fp})$. Each campaign cell contributes one
  coefficient; the median, the interquartile range [Q1--Q3, the middle half of the cells] and
  the full range (min--max) are taken over cells with $n\ge50$, and cells below that are
  omitted. $n_{fp}=5$ never reaches $50$ in any campaign
  (maximum $19$) and is therefore absent.}
  \label{tab:dataset}
  \setlength{\tabcolsep}{5pt}
  \renewcommand{\arraystretch}{1.2}
  \small
  \begin{tabular}{c c r c cc cc}
\toprule
& & & & \multicolumn{2}{c}{Spearman $\rho$} & \multicolumn{2}{c}{Pearson $r$} \\
\cmidrule(lr){5-6}\cmidrule(lr){7-8}
$n_{fp}$ & cells & $\sum n$ & $n$ per cell
  & median [IQR] & (range) & median [IQR] & (range) \\
\midrule
    2 & 16 & 4089 & 59--483 & $0.776$ [$0.765$--$0.802$] & ($0.683$--$0.839$) & $0.789$ [$0.776$--$0.810$] & ($0.709$--$0.845$) \\
    3 & 43 & 14455 & 83--661 & $0.717$ [$0.670$--$0.769$] & ($0.430$--$0.813$) & $0.722$ [$0.676$--$0.774$] & ($0.442$--$0.836$) \\
    4 & 33 & 3934 & 53--201 & $0.684$ [$0.592$--$0.735$] & ($0.445$--$0.850$) & $0.668$ [$0.604$--$0.714$] & ($0.403$--$0.790$) \\
    \midrule
    all & 92 & 22478 & 53--661 & $0.729$ [$0.642$--$0.774$] & ($0.430$--$0.850$) & $0.717$ [$0.650$--$0.777$] & ($0.403$--$0.845$) \\
\bottomrule
\end{tabular}

\end{table}

\begin{figure}[htbp]
  \centering
  \includegraphics[width=\linewidth]{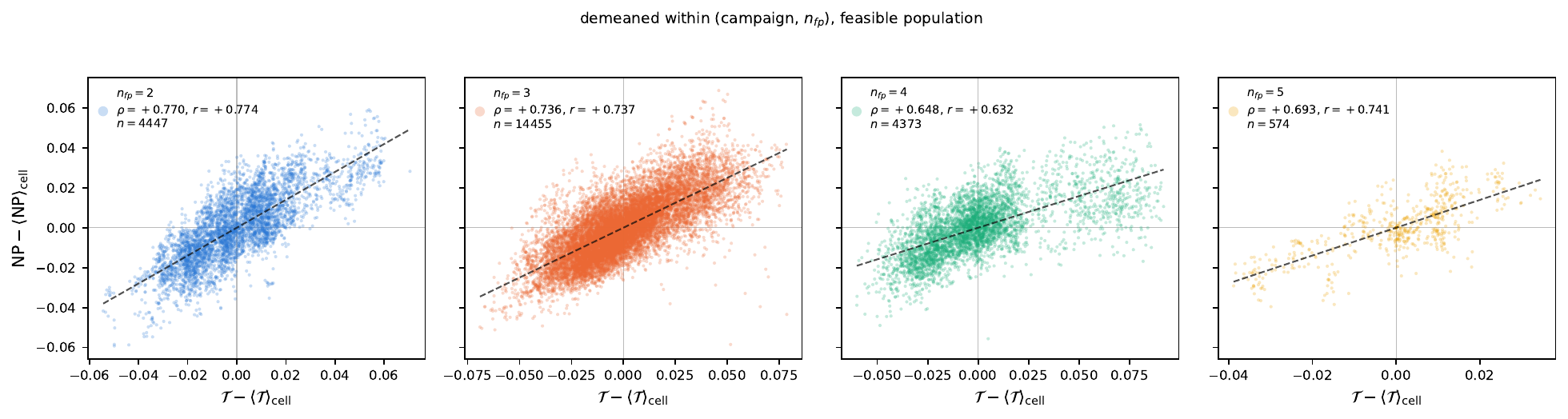}
  \caption{Mean coil non-planarity against $\mathcal{T}$, one panel per field-period count,
  pooled over all campaigns. Both quantities are demeaned within each
  $(\text{campaign},n_{fp})$ cell. Dashed lines are least-squares fits, shown to guide the
  eye; the quoted statistic is the Spearman coefficient.}
  \label{fig:scatter}
\end{figure}

Every one of the $32$ cells is positive, and all lie between $0.52$ and $0.85$. Pearson and
Spearman agree closely throughout, so the relation is close to linear rather than merely
monotonic. The spread is narrowest at $n_{fp}=2$ and widest at $n_{fp}=4$, where the cells are
smallest (Fig.~\ref{fig:scatter}). The middle half of the cells, however, spans only $0.02$,
$0.04$ and $0.13$ at
$n_{fp}=2$, $3$ and $4$, against full ranges of $0.08$, $0.27$ and $0.33$: in each row the
dispersion is carried by one or two outlying cells, and the typical cell agrees with its
neighbours far more closely than the range alone would suggest.

Table~\ref{tab:pooled} takes the complementary view: instead of one coefficient per cell, it
aggregates the cells of a field-period class and reports a single pair of coefficients for each.
When aggregating multiple cells together, we subtract each cell's mean from both predictor and 
target quantities. This places every cell's centroid at the origin, 
which removes the between-cell term of
\begin{equation}
  \operatorname{Cov}(x,y) = \mathbb{E}\!\left[\operatorname{Cov}(x,y \mid \text{cell})\right]
  + \operatorname{Cov}\!\left(\mathbb{E}[x \mid \text{cell}],\, \mathbb{E}[y \mid \text{cell}]\right),
  \label{eq:covdecomp}
\end{equation}
so the pooled coefficient measures only variation at fixed constraint set and cannot inherit a
trend from how the clouds happen to be arranged. Pooling is also the
only way to quantify $n_{fp}=5$ at all, since no single campaign reaches $50$ feasible
configurations there.

\begin{table}[htbp]
\centering
\caption{Across-campaign pooled correlation between $\mathcal{T}$ and mean coil non-planarity,
one coefficient pair per field-period count, demeaned within $(\text{campaign},n_{fp})$ before
pooling. `boundaries' counts the distinct plasma boundaries behind the $n$ coil sets: a
boundary recurs once per campaign in which it is feasible, so the effective sample is smaller
than $n$ and the intervals are optimistic --- severely so at $n_{fp}=5$, where $574$ coil sets
rest on $27$ boundaries.}
\label{tab:pooled}
\setlength{\tabcolsep}{5pt}
\renewcommand{\arraystretch}{1.2}
\small
\begin{tabular}{c rrrr cc cc}
\toprule
& & & & & \multicolumn{2}{c}{Spearman $\rho$} & \multicolumn{2}{c}{Pearson $r$} \\
\cmidrule(lr){6-7}\cmidrule(lr){8-9}
$n_{fp}$ & campaigns & cells & $n$ & boundaries & value & 95\% CI & value & 95\% CI \\
\midrule
    $2$ & 43 & 43 & 4447 & 535 & $0.770$ & $[0.758,\,0.782]$ & $0.774$ & $[0.762,\,0.786]$ \\
    $3$ & 43 & 43 & 14455 & 963 & $0.736$ & $[0.729,\,0.743]$ & $0.737$ & $[0.730,\,0.744]$ \\
    $4$ & 54 & 54 & 4373 & 346 & $0.648$ & $[0.630,\,0.665]$ & $0.632$ & $[0.614,\,0.650]$ \\
    $5$ & 54 & 54 & 574 & 27 & $0.693$ & $[0.648,\,0.733]$ & $0.741$ & $[0.701,\,0.775]$ \\
    \midrule
    all & 54 & 194 & 23849 & 1871 & $0.723$ & $[0.717,\,0.729]$ & $0.709$ & $[0.703,\,0.715]$ \\
\bottomrule
\end{tabular}

\end{table}

Read against Table~\ref{tab:dataset} the two agree where they overlap, as they must. The
pooled $n_{fp}=5$ coefficient, $\rho=0.68$, sits between the $n_{fp}=3$ and $n_{fp}=4$ values
rather than apart from them, which is the most that can be said for a field-period count whose
positive-well population is this thin upstream (Table~\ref{tab:nfp-population}).

Figure~\ref{fig:scatter} shows the underlying data. Both axes are demeaned within each
$(\text{campaign},n_{fp})$ cell before pooling, so the clouds display the same within-cell
variation the table summarises, with the offsets between constraint sets removed. The
relation is visibly linear and of uniform scatter across the range, with no threshold or
saturation. Pooled over all cells it gives $\rho = 0.723$ on $23849$ coil sets, drawn from
$1871$ distinct boundaries.
Figure~\ref{fig:coils3d} shows the same contrast on two individual members of this
population.

\subsection{Correlations of surface geometry and other magnetic predictors}
\label{sec:other-predictors}

The procedure of Sec.~\ref{sec:dataset-results} is applied unchanged to any other scalar
derived from the equilibrium and listed in Table~\ref{tab:predictors}. 
Population, filter, demeaning and cell cut are identical to Table~\ref{tab:dataset}; 
only the predictor changes.
The full tables, stratified by $n_{fp}$, are given in Appendix~\ref{app:predictors}.

The outcome is that $\mathcal{T}$ is the strongest of the sixteen and the physics scalars are
all weak. The best of them, the edge mirror ratio, reaches only median $\rho = 0.333$, with
$q_i$ at $0.229$ and the vacuum well, the axis mirror ratio and $L_{\nabla B}$ all below
$0.18$ in absolute value; for the two rotational transforms the interquartile range straddles
zero, so neither orders the population at all. Coil non-planarity here is not set by the
physics targets but by the boundary geometry. The one serious rival is the field-line normal
curvature, at median $\rho = 0.650$ against $0.729$ for $\mathcal{T}$; the two are not
independent as shown by \eqref{eq:mohr} and separating them is left to future work.

\begin{figure}[htbp]
  \centering
  \includegraphics[width=0.86\linewidth]{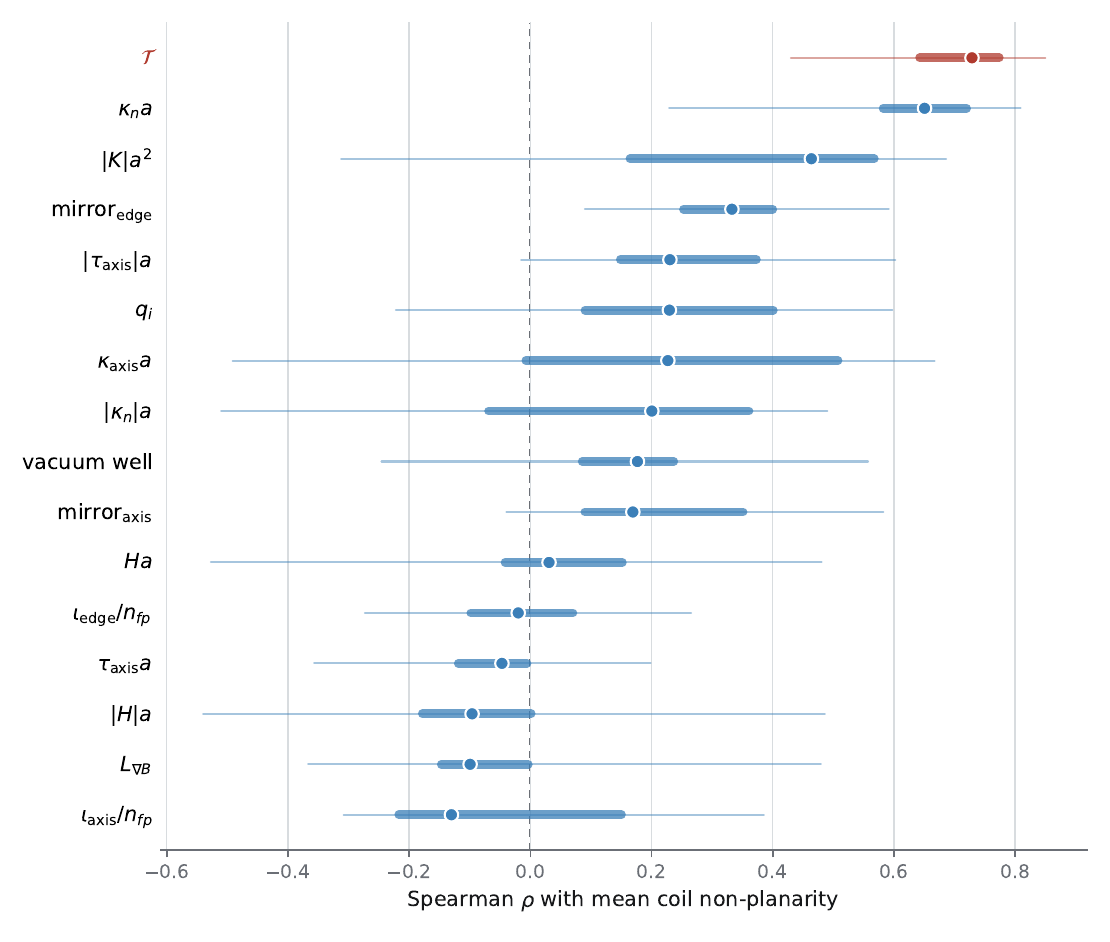}
  \caption{Spearman coefficient between each candidate predictor and the mean coil
  non-planarity The dot is the median over cells, 
  the thick bar the interquartile range and the thin line the full
  range; these are the three numbers tabulated in Appendix~\ref{app:predictors}. 
  Both quantities are demeaned within each cell, so every coefficient measures
  variation at fixed constraint set and field-period count. $\mathcal{T}$ is marked in red.
  The physics scalars cluster near zero with interquartile ranges straddling it; only the
  field-line normal curvature $\kappa_n a$ comes close, which \eqref{eq:mohr} ties to
  $\taug$ at every point of the boundary.}
  \label{fig:predictor-forest}
\end{figure}

\begin{figure}[htbp]
  \centering
  \includegraphics[width=\linewidth]{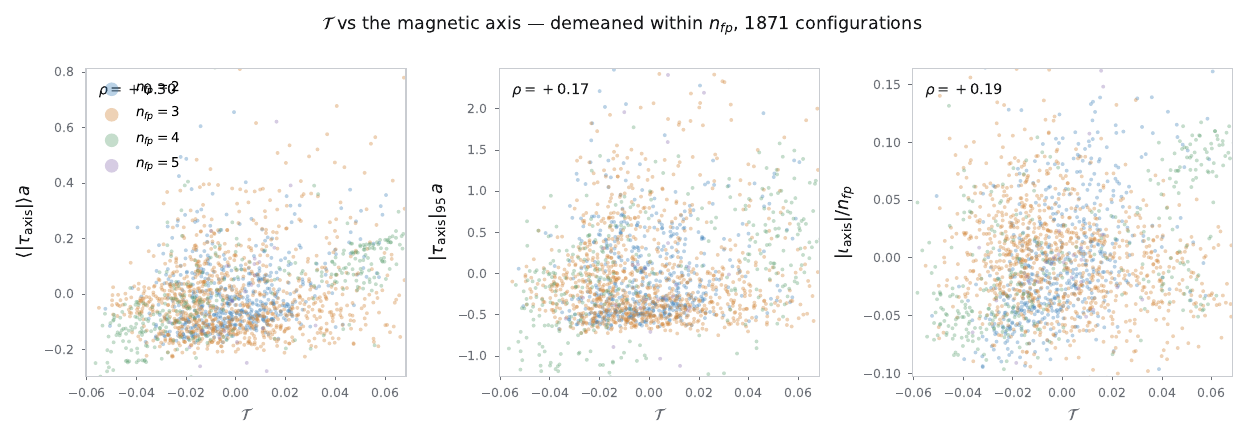}
  \caption{$\mathcal{T}$ against properties of the magnetic axis.}
  \label{fig:tg-axis}
\end{figure}

\begin{figure}[htbp]
  \centering
  \includegraphics[width=\linewidth]{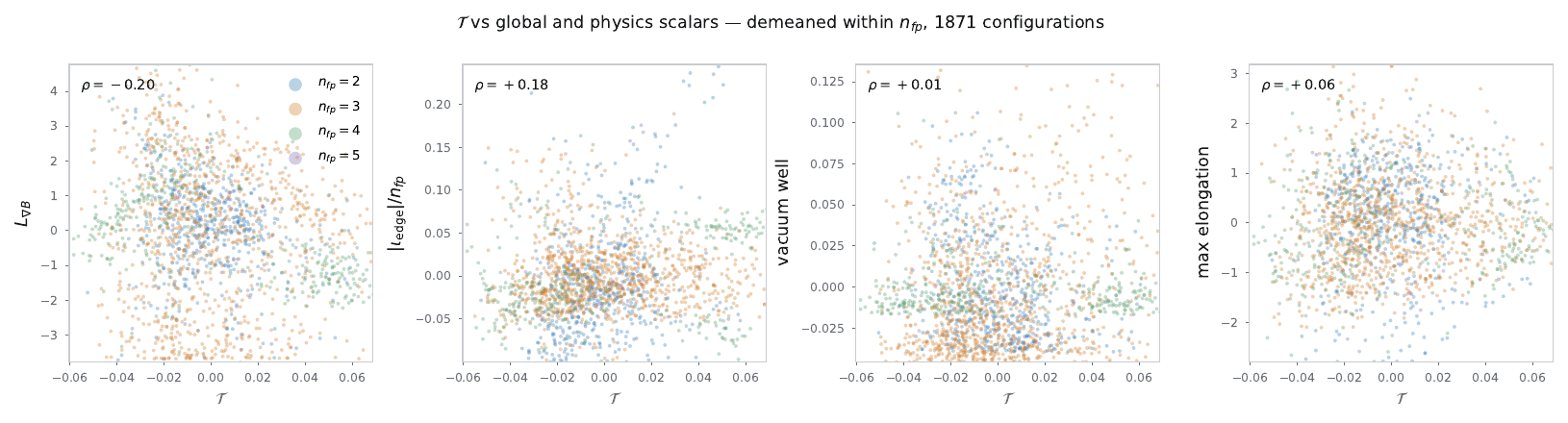}
  \caption{$\mathcal{T}$ against global and physics scalars.}
  \label{fig:tg-global}
\end{figure}

\begin{figure}[htbp]
  \centering
  \includegraphics[width=\linewidth]{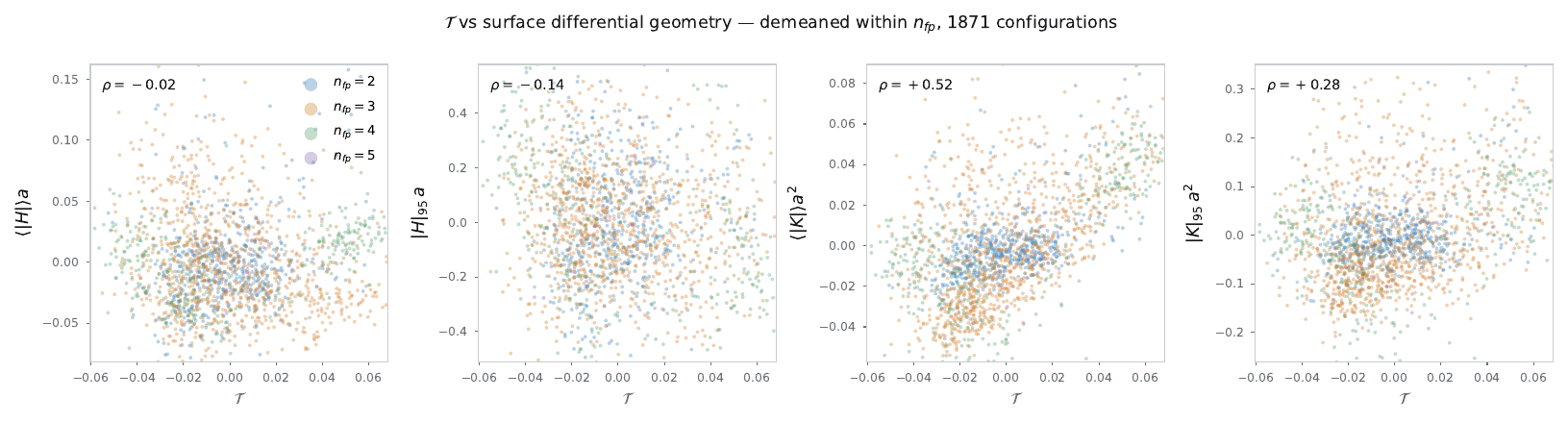}
  \caption{$\mathcal{T}$ against the surface differential geometry.}
  \label{fig:tg-surface}
\end{figure}

\begin{figure}[htbp]
  \centering
  \includegraphics[width=\linewidth]{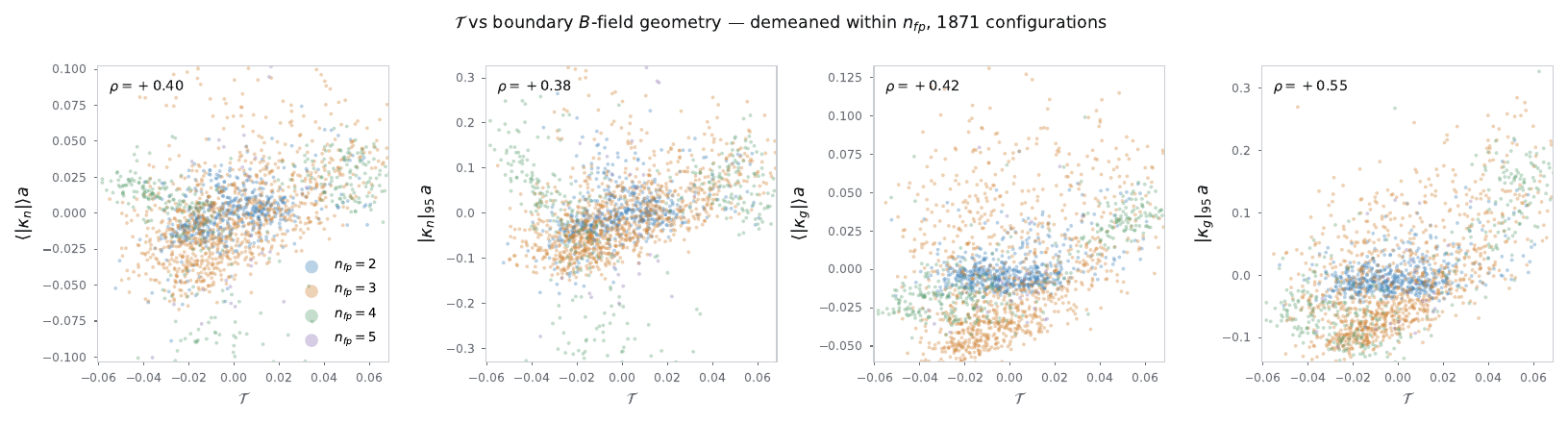}
  \caption{$\mathcal{T}$ against the other field-line curvatures. $\kappa_n$ is the predictor
  that comes closest to $\mathcal{T}$ against non-planarity (Sec.~\ref{sec:other-predictors}),
  so its correlation with $\mathcal{T}$ bounds how much of that is shared information.}
  \label{fig:tg-bfield}
\end{figure}

Whether $\mathcal{T}$ is merely a restatement of something already known is a separate
question from whether it predicts well, and Figs.~\ref{fig:tg-axis}--\ref{fig:tg-bfield}
address it directly: $\mathcal{T}$ against the candidate quantities themselves, grouped by
what they describe. Both axes are demeaned within each $n_{fp}$ population before the panels are
pooled, so the values shown are departures from the population mean and take both signs. The
strongest pooled association
is $\rho=0.55$, against $0.72$ for $\mathcal{T}$ with non-planarity: no quantity
considered here is a substitute for it. The closest relatives are the geodesic curvature of the
field line ($0.55$) and the Gaussian curvature ($0.52$), while the axis quantities and the
global scalars stay at or below $0.30$ and $0.20$ respectively. For $K$ this could be explained by \eqref{eq:mohr}. We note that the same relation involves
$H$, which nonetheless ranks last, so it does not by itself account for the ordering of the
two. Appendix~\ref{app:taug-pairs} gives the coefficients stratified by $n_{fp}$.

\subsection{Published configurations outside the dataset}
\label{sec:known}

We compare $\mathcal{T}$ with the coil non-planarity for five published vacuum
stellarator configurations outside the dataset: precise QI at $n_{fp}=2$ and $n_{fp}=3$
from~\cite{goodman}, precise QA and precise QH from~\cite{landremanpaul}, and the W7-X
standard vacuum configuration.

The five differ in field-period count, quasi-symmetry family and aspect ratio, so a single
coil layout cannot serve all of them. The three geometric requirements --- minimum
coil--surface distance, minimum coil--coil distance and maximum curvature --- are external
constraints set by, for example a reactor design, and are held common. 
The number of unique coils per half
period and the coil length are coupled to each other
through the length reference $2\pi(a+d_{cs})n_c$, and $n_c$ is tied to $n_{fp}$ through the
toroidal coil spacing $(\pi/n_{fp})R/(n_c a)$. Imposing a common $n_c$ would therefore not 
place the configurations on an equal footing but would give the low-$n_{fp}$ cases a coarser 
coil set than the others. For each configuration we instead take the smallest $n_c$ and length 
budget that reconstruct the field to the required accuracy while respecting the three common
constraints, found through a coarse grid scan described in Appendix~\ref{app:scan}. 
Table~\ref{tab:published} reports the result.

\begin{table}[htbp]
  \centering
  \caption{Stage-two filament coil sets for the five published configurations, ordered by
  $\mathcal{T}$. All five satisfy the same geometric requirements, $d_{cs}/a\ge0.70$,
  $d_{cc}/a\ge0.40$ and $k_{\max}a\le3.0$, and each is given the smallest $n_c$ and length
  budget that reconstruct the field within them. The normal field error $\Bnorm$ is below the 
  $5\times10^{-3}$ threshold for all five; the coil length is given per minor radius. 
  Pooled over all five, ordering by $\mathcal{T}$ reproduces the ordering of the mean 
  non-planarity with two inversions, precise QI $n_{fp}=2$ against precise QH $n_{fp}=4$
  and precise QI $n_{fp}=3$ against W7-X $n_{fp}=5$.}
  \label{tab:published}
  \setlength{\tabcolsep}{8pt}
  \renewcommand{\arraystretch}{1.2}
    \begin{tabular}{l c c c c c c}
    \toprule
    configuration & $n_{fp}$ & $\mathcal{T}$ & $\Bnorm$ & mean NP & $n_c$ & $L/a$ \\
    \midrule
    precise QI, $n_{fp}{=}2$ & 2 & 0.0694 & $1.3\times10^{-3}$ & 0.0758 & 14 & 231.8 \\
    precise QH               & 4 & 0.0774 & $4.8\times10^{-3}$ & 0.0676 &  6 &  83.3 \\
    precise QA               & 2 & 0.0885 & $1.5\times10^{-3}$ & 0.1001 &  7 & 119.6 \\
    precise QI, $n_{fp}{=}3$ & 3 & 0.0947 & $4.9\times10^{-3}$ & 0.1262 &  6 & 109.0 \\
    W7-X standard            & 5 & 0.1321 & $3.9\times10^{-3}$ & 0.1148 &  5 &  74.8 \\
    \bottomrule
  \end{tabular}

\end{table}

Ordering by $\mathcal{T}$ reproduces the ordering of the mean non-planarity with two inversions.
Both inversions in Table~\ref{tab:published} are between configurations whose field-period
counts differ by two. That is the least controlled comparison available here: $n_{fp}$ fixes
the toroidal spacing of the coils through $(\pi/n_{fp})R/(n_c a)$, so configurations far
apart in $n_{fp}$ are not presented with the same coil-design problem even when the
geometric requirements and the budget rule are identical. The best controlled comparison is therefore between two
configurations of the same $n_{fp}$, and the next best between counts differing by one. Only
one pair in this set shares an $n_{fp}$, so Table~\ref{tab:nfpblocks} groups the
configurations into the three overlapping blocks for which one of these two holds. Within
every block the mean non-planarity follows $\mathcal{T}$ without inversion.

\begin{table}[htbp]
  \centering
  \caption{The coil sets of Table~\ref{tab:published}, grouped into the three blocks of
  configurations whose field-period counts differ by at most one, each block ordered by
  $\mathcal{T}$. Precise QI $n_{fp}=2$ and precise QA share a field-period count, which makes
  theirs the only fully controlled comparison in the set. Precise QI $n_{fp}=3$ and precise
  QH each belong to two blocks and are listed twice; the coil sets are those of
  Table~\ref{tab:published} and are not re-optimised. Within every block the mean
  non-planarity follows $\mathcal{T}$ without inversion.}
  \label{tab:nfpblocks}
  \setlength{\tabcolsep}{8pt}
  \renewcommand{\arraystretch}{1.2}
    \begin{tabular}{l c c c c c c}
    \toprule
    configuration & $n_{fp}$ & $\mathcal{T}$ & $\Bnorm$ & mean NP & $n_c$ & $L/a$ \\
    \midrule
    \multicolumn{7}{l}{\textit{$n_{fp}=2$ and $n_{fp}=3$}} \\
    precise QI, $n_{fp}{=}2$ & 2 & 0.0694 & $1.3\times10^{-3}$ & 0.0758 & 14 & 231.8 \\
    precise QA               & 2 & 0.0885 & $1.5\times10^{-3}$ & 0.1001 &  7 & 119.6 \\
    precise QI, $n_{fp}{=}3$ & 3 & 0.0947 & $4.9\times10^{-3}$ & 0.1262 &  6 & 109.0 \\
    \addlinespace
    \multicolumn{7}{l}{\textit{$n_{fp}=3$ and $n_{fp}=4$}} \\
    precise QH               & 4 & 0.0774 & $4.8\times10^{-3}$ & 0.0676 &  6 &  83.3 \\
    precise QI, $n_{fp}{=}3$ & 3 & 0.0947 & $4.9\times10^{-3}$ & 0.1262 &  6 & 109.0 \\
    \addlinespace
    \multicolumn{7}{l}{\textit{$n_{fp}=4$ and $n_{fp}=5$}} \\
    precise QH               & 4 & 0.0774 & $4.8\times10^{-3}$ & 0.0676 &  6 &  83.3 \\
    W7-X standard            & 5 & 0.1321 & $3.9\times10^{-3}$ & 0.1148 &  5 &  74.8 \\
    \bottomrule
  \end{tabular}

\end{table}

The non-planarity obtained is a property of the configuration rather than of the particular
requirements imposed on the coils. Repeating the optimisation with the coil--surface
requirement relaxed to $d_{cs}/a\ge0.60$, the other two unchanged, gives
Table~\ref{tab:published-dcs06}: the same $n_c$ for every configuration, budgets within a
few per cent of those in Table~\ref{tab:published}, and mean non-planarities that differ by
at most $2.2\%$. The ordering, and the single inversion between the two configurations two
field periods apart, are unchanged. Precise QI $n_{fp}=2$ has no solution satisfying the
relaxed set at any budget we tried and is absent.

\begin{table}[htbp]
  \centering
  \caption{As Table~\ref{tab:published}, with the coil--surface requirement relaxed to
  $d_{cs}/a\ge0.60$ and $d_{cc}/a\ge0.40$, $k_{\max}a\le3.0$ unchanged. Precise QI
  $n_{fp}=2$ admits no solution under this set.}
  \label{tab:published-dcs06}
  \setlength{\tabcolsep}{8pt}
  \renewcommand{\arraystretch}{1.2}
    \begin{tabular}{l c c c c c c}
    \toprule
    configuration & $n_{fp}$ & $\mathcal{T}$ & $\Bnorm$ & mean NP & $n_c$ & $L/a$ \\
    \midrule
    precise QH               & 4 & 0.0774 & $4.3\times10^{-3}$ & 0.0687 & 6 &  84.4 \\
    precise QA               & 2 & 0.0885 & $2.1\times10^{-3}$ & 0.0979 & 7 & 116.1 \\
    precise QI, $n_{fp}{=}3$ & 3 & 0.0947 & $4.2\times10^{-3}$ & 0.1287 & 6 & 111.6 \\
    W7-X standard            & 5 & 0.1321 & $3.5\times10^{-3}$ & 0.1150 & 5 &  75.4 \\
    \bottomrule
  \end{tabular}

\end{table}

\begin{figure}[htbp]
  \centering
  \includegraphics[width=\linewidth]{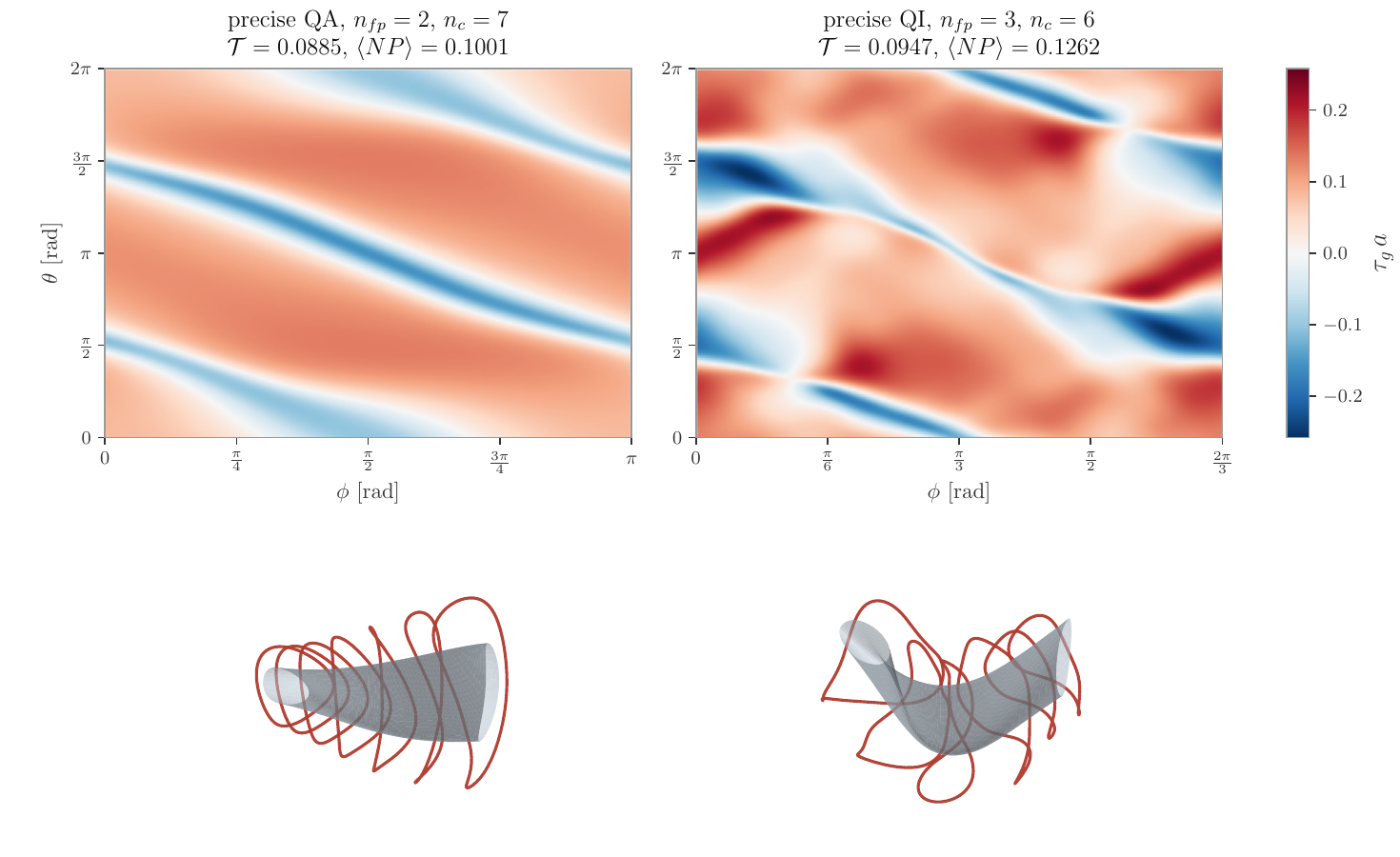}
  \caption{Geodesic torsion $\taugB a$ over one field period (top) and the boundary with its
  stage-two coils over one half field period (bottom), for precise QA (left) and precise QI
  $n_{fp}=3$ (right). The two have nearly the same $\mathcal{T}$ but different distributions
  of it; the comparison is qualitative.}
  \label{fig:stage2geom}
\end{figure}

Figure~\ref{fig:stage2geom} shows the two configurations that sit closest in $\mathcal{T}$,
precise QA and precise QI $n_{fp}=3$, as the geodesic torsion on the boundary and the coil
set obtained for it.

\section{Conclusions and future work}
\label{sec:conclusions}

The Frenet torsion of the Rodr\'iguez--Sengupta current contours --- the quantity that
vanishes exactly when the ideal coil is planar --- can be written entirely in terms of the
flux surface and the magnetic field on it, Eq.~\eqref{eq:tauK}. Of its two terms, one is the
geodesic torsion of the field lines: simple to compute from the surface geometry, bounded,
frame-independent, and defined everywhere on the surface. The other carries the coil curvature
in its denominator and is not well defined where that curvature vanishes.

Taking the area-weighted surface average of the first term alone, we find that it predicts the
non-planarity of optimised filament coils at Spearman $\rho=0.52$--$0.85$ across every
field-period class and every constraint set in the dataset, and that the ranking it induces
reproduces the ordering of five published configurations outside the dataset.

Two limitations should be stated. The first is sample size: the population is thin at
$n_{fp}=5$, where no single campaign reaches fifty feasible configurations and the pooled
coefficient rests on $25$ distinct boundaries. The second is the composition of the dataset
itself. Its boundaries admit comparatively large mirror ratios, elongations and rotational
transforms, and they carry poloidal modes
only up to $m=4$. Both restrict the family of shapes sampled, so we cannot rule out that the
trends established here are driven by the particular covariation of features present in the
original stage-one equilibria. Extending the study to a broader and more diverse range of
equilibria is left for future work.

The geodesic torsion is also only the part of the problem that is easy to compute. The second
term of \eqref{eq:tauK} must account for something in at least some cases, and we have not
attempted to say when. That question is left open here and is worth pursuing.

A natural next step is to use $\mathcal{T}$ as a stage-one objective. We have established
trends for this one scalar, but whether this reduction of the two-dimensional $\taugB$ field
is the right quantity to optimise against is a question for a dedicated study. The planarity
condition $\tau^{(K)}=0$ suggests two distinct penalties: minimising $\langle|\taugB|\rangle$,
which suppresses the surface-imposed term alone, or penalising the residual of the full
condition \eqref{eq:tauK}, which targets planarity itself at the cost of the ill-conditioned
second term. Which of the two is the better objective is an open question.

\section*{Acknowledgements}
We thank Eduardo Rodr\'iguez, Pedro F. Gil, J\k{e}drzej Walkowiak, Gabriel G. Plunk
and Michael Drevlak for useful and insightful discussions.

\appendix
\section{Computing the geodesic torsion}
\label{app:taug-numerics}

For a field line with unit tangent $\bhat = \Bvec/|\Bvec|$ on a surface with unit normal $\nhat$, the geodesic torsion is
the second fundamental form evaluated on $\bhat$ and the co-normal $\hat{\boldsymbol g} =
\nhat \times \bhat$,
\begin{equation}
  \taug \;=\; \mathrm{II}\!\left(\bhat,\, \hat{\boldsymbol g}\right).
  \label{eq:taug-II}
\end{equation}
Expanding each tangent vector in the surface basis $(\boldsymbol r_\phi, \boldsymbol
r_\theta)$ as $\bhat = b^\phi \boldsymbol r_\phi + b^\theta \boldsymbol r_\theta$ gives
\begin{equation}
  \taug \;=\; L\,b^\phi g^\phi \;+\; M\left(b^\phi g^\theta + b^\theta g^\phi\right)
            \;+\; N\,b^\theta g^\theta ,
  \label{eq:taug-LMN}
\end{equation}
with $L$, $M$, $N$ the coefficients of the second fundamental form. The contravariant
components follow from inverting the first fundamental form,
\begin{equation}
  b^\phi = \frac{G\,(\bhat\cdot\boldsymbol r_\phi) - F\,(\bhat\cdot\boldsymbol r_\theta)}
                {EG - F^2} ,
  \qquad
  b^\theta = \frac{E\,(\bhat\cdot\boldsymbol r_\theta) - F\,(\bhat\cdot\boldsymbol r_\phi)}
                  {EG - F^2} ,
  \label{eq:contravariant}
\end{equation}
and likewise for $\hat{\boldsymbol g}$. The inversion is always well posed: by Lagrange's
identity $EG - F^2 = |\boldsymbol r_\phi \times \boldsymbol r_\theta|^2$, which is the square
of the area element and is therefore strictly positive wherever the parametrisation is regular.
It could vanish only where $\boldsymbol r_\phi$ and $\boldsymbol r_\theta$ become linearly
dependent, which on a flux surface happens at the magnetic axis and not on the boundary.

Every quantity in \eqref{eq:taug-LMN}--\eqref{eq:contravariant} is supplied by
\textsc{simsopt}~\cite{simsopt}, on one grid, with no geometry reimplemented:

\begin{center}
\setlength{\tabcolsep}{9pt}
\renewcommand{\arraystretch}{1.25}
\begin{tabular}{l l}
  \toprule
  quantity & \textsc{simsopt} call \\
  \midrule
  boundary surface                         & \texttt{SurfaceRZFourier.from\_wout(wout, s=1)} \\
  $\boldsymbol r_\phi$, $\boldsymbol r_\theta$ & \texttt{surface.gammadash1()}, \texttt{surface.gammadash2()} \\
  $\nhat$                                  & \texttt{surface.unitnormal()} \\
  $E$, $F$, $G$                            & \texttt{surface.first\_fund\_form()} \\
  $L$, $M$, $N$                            & \texttt{surface.second\_fund\_form()} \\
  $\Bvec$ on the same points               & \texttt{vmec\_compute\_geometry(vmec, s, theta, phi)} \\
  \bottomrule
\end{tabular}
\end{center}

\noindent
Equations~\eqref{eq:taug-II}--\eqref{eq:contravariant} are pointwise algebra on the surface.
No derivative of the field is taken numerically, so $\taugB$ carries no grid-convergence
question of its own and is defined at every point of the boundary, including umbilics. This is
the property that distinguishes it from the second term of \eqref{eq:tauK}, and the reason the
area-weighted average \eqref{eq:T} is well posed.

\section{Choosing \texorpdfstring{$n_c$}{nc} and the length budget}
\label{app:scan}

The coil sets of Table~\ref{tab:published} are the outcome of a coarse scan over two
quantities, the number of unique coils per half period $n_c$ and the length target ratio
$\ell$ of Eq.~\eqref{eq:jl}. The three geometric requirements are held fixed throughout.
For each configuration we report the smallest budget at which the optimiser meets all three
requirements outright and the normal field error \eqref{eq:fielderr} falls below
$\Bnorm=5\times10^{-3}$; in every reported
case the length constraint is active, so the coil set is as short as that requirement set
permits.

The starting $n_c$ is the value for which the toroidal coil spacing $(\pi/n_{fp})R/(n_c a)$
matches that of the W7-X coil set, and $\ell$ is then raised in steps of $0.05$ to $0.10$
from a value at which $\Bnorm$ is still above threshold. Where no $\ell$ gives a
feasible point, $n_c$ is changed and the scan repeated. Table~\ref{tab:scan} summarises what
was covered.

\begin{table}[htbp]
  \centering
  \caption{The scan behind Table~\ref{tab:published}: values of $n_c$ tried, the range of
  length fractions $\ell$ covered, the number of optimisations run, and the requirement that
  fails below the reported point. Twenty-five optimisations in total.}
  \label{tab:scan}
  \setlength{\tabcolsep}{7pt}
  \renewcommand{\arraystretch}{1.2}
    \begin{tabular}{l c c c l}
    \toprule
    configuration & $n_c$ tried & $\ell$ range & runs & binding failure below the reported point \\
    \midrule
    precise QI, $n_{fp}{=}2$ & 12, 14   & 0.35--0.55 & 6 & $d_{cs}$, by $5.9$--$12.5\%$ at $n_c=12$ \\
    precise QH               & 6        & 0.30--0.60 & 4 & none; feasible at the smallest $\ell$ tried \\
    precise QA               & 7        & 0.40--0.60 & 4 & $d_{cs}$, by $1.0$--$1.5\%$ \\
    precise QI, $n_{fp}{=}3$ & 6, 7, 8  & 0.45--0.90 & 9 & $d_{cs}$ at $n_c=7$, $d_{cc}$ at $n_c=8$ \\
    W7-X standard            & 5        & 0.35--0.40 & 2 & $\Bnorm$, $6.3\times10^{-3}$ \\
    \bottomrule
  \end{tabular}

\end{table}

Two configurations needed an $n_c$ away from the spacing estimate, and they fail in opposite
ways. Precise QI $n_{fp}=3$ at $n_c=7$ misses the coil--surface requirement at every budget
scanned, closing from $40.4\%$ to $0.8\%$ as $\ell$ goes from $0.45$ to $0.60$ but not
reaching it; raising $n_c$ to $8$ buys the clearance and then violates the coil--coil
requirement by $3.9\%$ instead, the two distances closing on each other as the coils are
packed more tightly. Lowering $n_c$ to $6$ resolves both, and does so at a shorter total
length, $109.0a$ against the $119.6a$ of the infeasible $n_c=7$ point. Precise QI $n_{fp}=2$
at $n_c=12$ misses the coil--surface requirement by $5.9\%$ even at the longest budget
scanned, and lengthening does not close the gap; at $n_c=14$ the miss falls to $3.5\%$ at
$\ell=0.45$ and the requirement is reached at $\ell=0.55$, where the achieved clearance sits
on the bound to within $0.05\%$. This configuration carries by far the largest budget of the
five, $231.8a$ against $74.8$--$119.6a$ for the others.

Precise QH is feasible at the smallest $\ell$ scanned, so its entry is an upper bound on the
minimum rather than a located one. The configurations of Table~\ref{tab:published-dcs06}
were scanned in the same way and over comparable ranges, twenty-five optimisations in total;
precise QI $n_{fp}=2$ was carried to $\ell=0.70$ at $n_c=12$ without reaching the relaxed
requirement, and is absent from that table for this reason.

\section{Correlation of the other candidate predictors}
\label{app:predictors}

Tables~\ref{tab:pred-nfp2}--\ref{tab:pred-nfpall} give the full tables referred to in
Sec.~\ref{sec:other-predictors}: Spearman and Pearson coefficients between each candidate
predictor and the mean coil non-planarity, on the same population, filter, demeaning and
cell cut as Table~\ref{tab:dataset}. Surface fields are reduced by an area-weighted mean and
made dimensionless with the minor radius, as $\mathcal{T}$ is; where the sign of a field
carries meaning, both the signed and the absolute reduction are listed, since which of the two
correlates is itself informative. Rows are ordered by $|$median $\rho|$ pooled over all cells,
and $\mathcal{T}$ is repeated in each table as the reference.

\begin{table}[htbp]
  \centering
  \caption{Correlation between each predictor and mean coil non-planarity, $n_{fp}=2$. Feasible population, demeaned within $(\text{campaign},n_{fp})$; 16 campaign cells with $n\ge50$, $\sum n=4089$, 59--483 per cell. Each entry is the median [Q1, Q3] (min--max) over those cells. Surface fields are reduced by an area-weighted mean and made dimensionless with the minor radius; both signed and absolute reductions are given where the sign carries meaning. Rows are ordered by $|$median $\rho|$ pooled over all cells.}
  \label{tab:pred-nfp2}
  \setlength{\tabcolsep}{6pt}
  \renewcommand{\arraystretch}{1.15}
  \small
  \begin{tabular}{lcc}
    \toprule
    predictor & Spearman $\rho$ & Pearson $r$ \\
    \midrule
    $\mathcal{T}$ & $+0.776$ [$+0.765$, $+0.802$] ($+0.683$--$+0.839$) & $+0.789$ [$+0.776$, $+0.810$] ($+0.709$--$+0.845$) \\
    $\kappa_n a$ & $+0.744$ [$+0.719$, $+0.779$] ($+0.656$--$+0.809$) & $+0.708$ [$+0.664$, $+0.736$] ($+0.642$--$+0.762$) \\
    $|K| a^2$ & $+0.556$ [$+0.526$, $+0.567$] ($+0.469$--$+0.596$) & $+0.525$ [$+0.503$, $+0.535$] ($+0.421$--$+0.580$) \\
    mirror$_{\rm edge}$ & $+0.182$ [$+0.153$, $+0.307$] ($+0.090$--$+0.407$) & $+0.295$ [$+0.242$, $+0.386$] ($+0.166$--$+0.468$) \\
    $|\tau_{\rm axis}| a$ & $+0.228$ [$+0.159$, $+0.304$] ($+0.055$--$+0.382$) & $+0.198$ [$+0.135$, $+0.275$] ($+0.080$--$+0.379$) \\
    $q_i$ & $+0.098$ [$+0.007$, $+0.139$] ($-0.072$--$+0.171$) & $+0.109$ [$+0.088$, $+0.155$] ($+0.078$--$+0.216$) \\
    $\kappa_{\rm axis} a$ & $+0.214$ [$+0.180$, $+0.223$] ($+0.054$--$+0.255$) & $+0.323$ [$+0.267$, $+0.344$] ($+0.240$--$+0.424$) \\
    $|\kappa_n| a$ & $+0.312$ [$+0.226$, $+0.340$] ($+0.053$--$+0.386$) & $+0.354$ [$+0.302$, $+0.375$] ($+0.140$--$+0.399$) \\
    vacuum well & $-0.153$ [$-0.198$, $-0.056$] ($-0.246$--$+0.135$) & $-0.147$ [$-0.192$, $-0.060$] ($-0.232$--$+0.235$) \\
    mirror$_{\rm axis}$ & $+0.154$ [$+0.113$, $+0.205$] ($+0.057$--$+0.292$) & $+0.202$ [$+0.145$, $+0.285$] ($+0.071$--$+0.306$) \\
    $\iota_{\rm axis}/n_{fp}$ & $+0.158$ [$+0.054$, $+0.194$] ($-0.195$--$+0.226$) & $-0.004$ [$-0.107$, $+0.039$] ($-0.265$--$+0.104$) \\
    $L_{\nabla B}$ & $-0.247$ [$-0.297$, $-0.216$] ($-0.367$--$-0.193$) & $-0.293$ [$-0.307$, $-0.273$] ($-0.367$--$-0.261$) \\
    $|H| a$ & $+0.338$ [$+0.319$, $+0.384$] ($+0.286$--$+0.486$) & $+0.303$ [$+0.287$, $+0.355$] ($+0.257$--$+0.488$) \\
    $\tau_{\rm axis} a$ & $-0.043$ [$-0.096$, $-0.010$] ($-0.201$--$+0.059$) & $-0.067$ [$-0.133$, $-0.040$] ($-0.393$--$+0.011$) \\
    $H a$ & $-0.409$ [$-0.431$, $-0.379$] ($-0.527$--$-0.305$) & $-0.383$ [$-0.438$, $-0.360$] ($-0.542$--$-0.307$) \\
    $\iota_{\rm edge}/n_{fp}$ & $-0.000$ [$-0.043$, $+0.059$] ($-0.177$--$+0.126$) & $-0.044$ [$-0.131$, $+0.009$] ($-0.274$--$+0.080$) \\
    \bottomrule
  \end{tabular}
\end{table}

\begin{table}[htbp]
  \centering
  \caption{Correlation between each predictor and mean coil non-planarity, $n_{fp}=3$. Feasible population, demeaned within $(\text{campaign},n_{fp})$; 43 campaign cells with $n\ge50$, $\sum n=14455$, 83--661 per cell. Each entry is the median [Q1, Q3] (min--max) over those cells. Surface fields are reduced by an area-weighted mean and made dimensionless with the minor radius; both signed and absolute reductions are given where the sign carries meaning. Rows are ordered by $|$median $\rho|$ pooled over all cells.}
  \label{tab:pred-nfp3}
  \setlength{\tabcolsep}{6pt}
  \renewcommand{\arraystretch}{1.15}
  \small
  \begin{tabular}{lcc}
    \toprule
    predictor & Spearman $\rho$ & Pearson $r$ \\
    \midrule
    $\mathcal{T}$ & $+0.717$ [$+0.670$, $+0.769$] ($+0.430$--$+0.813$) & $+0.722$ [$+0.676$, $+0.774$] ($+0.442$--$+0.836$) \\
    $\kappa_n a$ & $+0.664$ [$+0.611$, $+0.720$] ($+0.314$--$+0.760$) & $+0.674$ [$+0.611$, $+0.716$] ($+0.315$--$+0.833$) \\
    $|K| a^2$ & $+0.556$ [$+0.447$, $+0.628$] ($+0.338$--$+0.686$) & $+0.562$ [$+0.405$, $+0.614$] ($+0.295$--$+0.695$) \\
    mirror$_{\rm edge}$ & $+0.317$ [$+0.258$, $+0.367$] ($+0.130$--$+0.403$) & $+0.338$ [$+0.262$, $+0.377$] ($+0.109$--$+0.430$) \\
    $|\tau_{\rm axis}| a$ & $+0.156$ [$+0.113$, $+0.213$] ($+0.032$--$+0.272$) & $+0.170$ [$+0.103$, $+0.229$] ($-0.051$--$+0.288$) \\
    $q_i$ & $+0.415$ [$+0.301$, $+0.496$] ($+0.021$--$+0.597$) & $+0.366$ [$+0.265$, $+0.454$] ($+0.049$--$+0.588$) \\
    $\kappa_{\rm axis} a$ & $+0.525$ [$+0.411$, $+0.577$] ($+0.315$--$+0.667$) & $+0.472$ [$+0.372$, $+0.537$] ($+0.248$--$+0.642$) \\
    $|\kappa_n| a$ & $+0.354$ [$+0.198$, $+0.428$] ($+0.144$--$+0.490$) & $+0.348$ [$+0.286$, $+0.388$] ($+0.089$--$+0.479$) \\
    vacuum well & $+0.214$ [$+0.152$, $+0.250$] ($-0.039$--$+0.291$) & $+0.197$ [$+0.115$, $+0.254$] ($-0.135$--$+0.323$) \\
    mirror$_{\rm axis}$ & $+0.104$ [$+0.040$, $+0.155$] ($-0.039$--$+0.218$) & $+0.148$ [$+0.074$, $+0.178$] ($-0.034$--$+0.238$) \\
    $\iota_{\rm axis}/n_{fp}$ & $-0.223$ [$-0.260$, $-0.183$] ($-0.308$--$-0.087$) & $-0.221$ [$-0.245$, $-0.153$] ($-0.335$--$-0.066$) \\
    $L_{\nabla B}$ & $-0.104$ [$-0.125$, $-0.085$] ($-0.247$--$+0.001$) & $-0.091$ [$-0.115$, $-0.046$] ($-0.209$--$+0.010$) \\
    $|H| a$ & $-0.072$ [$-0.131$, $-0.025$] ($-0.271$--$+0.080$) & $-0.042$ [$-0.110$, $-0.005$] ($-0.188$--$+0.123$) \\
    $\tau_{\rm axis} a$ & $-0.015$ [$-0.047$, $+0.051$] ($-0.156$--$+0.199$) & $-0.048$ [$-0.091$, $-0.007$] ($-0.215$--$+0.138$) \\
    $H a$ & $+0.013$ [$-0.011$, $+0.049$] ($-0.082$--$+0.182$) & $-0.010$ [$-0.051$, $+0.022$] ($-0.139$--$+0.144$) \\
    $\iota_{\rm edge}/n_{fp}$ & $-0.076$ [$-0.104$, $-0.007$] ($-0.157$--$+0.151$) & $-0.107$ [$-0.149$, $-0.066$] ($-0.195$--$+0.089$) \\
    \bottomrule
  \end{tabular}
\end{table}

\begin{table}[htbp]
  \centering
  \caption{Correlation between each predictor and mean coil non-planarity, $n_{fp}=4$. Feasible population, demeaned within $(\text{campaign},n_{fp})$; 33 campaign cells with $n\ge50$, $\sum n=3934$, 53--201 per cell. Each entry is the median [Q1, Q3] (min--max) over those cells. Surface fields are reduced by an area-weighted mean and made dimensionless with the minor radius; both signed and absolute reductions are given where the sign carries meaning. Rows are ordered by $|$median $\rho|$ pooled over all cells.}
  \label{tab:pred-nfp4}
  \setlength{\tabcolsep}{6pt}
  \renewcommand{\arraystretch}{1.15}
  \small
  \begin{tabular}{lcc}
    \toprule
    predictor & Spearman $\rho$ & Pearson $r$ \\
    \midrule
    $\mathcal{T}$ & $+0.684$ [$+0.592$, $+0.735$] ($+0.445$--$+0.850$) & $+0.668$ [$+0.604$, $+0.714$] ($+0.403$--$+0.790$) \\
    $\kappa_n a$ & $+0.568$ [$+0.462$, $+0.646$] ($+0.229$--$+0.801$) & $+0.562$ [$+0.473$, $+0.619$] ($+0.342$--$+0.764$) \\
    $|K| a^2$ & $+0.074$ [$-0.072$, $+0.180$] ($-0.312$--$+0.465$) & $+0.182$ [$+0.007$, $+0.279$] ($-0.163$--$+0.489$) \\
    mirror$_{\rm edge}$ & $+0.428$ [$+0.352$, $+0.499$] ($+0.231$--$+0.591$) & $+0.496$ [$+0.403$, $+0.542$] ($+0.253$--$+0.655$) \\
    $|\tau_{\rm axis}| a$ & $+0.421$ [$+0.347$, $+0.454$] ($-0.015$--$+0.602$) & $+0.368$ [$+0.292$, $+0.421$] ($+0.016$--$+0.545$) \\
    $q_i$ & $+0.103$ [$-0.034$, $+0.189$] ($-0.222$--$+0.458$) & $+0.237$ [$+0.171$, $+0.357$] ($+0.009$--$+0.503$) \\
    $\kappa_{\rm axis} a$ & $-0.148$ [$-0.297$, $+0.006$] ($-0.491$--$+0.314$) & $-0.140$ [$-0.276$, $+0.028$] ($-0.458$--$+0.278$) \\
    $|\kappa_n| a$ & $-0.172$ [$-0.347$, $-0.034$] ($-0.510$--$+0.217$) & $-0.113$ [$-0.253$, $+0.020$] ($-0.435$--$+0.226$) \\
    vacuum well & $+0.181$ [$+0.127$, $+0.255$] ($-0.009$--$+0.557$) & $+0.165$ [$+0.097$, $+0.241$] ($-0.028$--$+0.481$) \\
    mirror$_{\rm axis}$ & $+0.412$ [$+0.342$, $+0.516$] ($+0.244$--$+0.583$) & $+0.455$ [$+0.406$, $+0.526$] ($+0.257$--$+0.615$) \\
    $\iota_{\rm axis}/n_{fp}$ & $+0.141$ [$+0.099$, $+0.192$] ($-0.206$--$+0.385$) & $+0.156$ [$+0.095$, $+0.212$] ($-0.164$--$+0.406$) \\
    $L_{\nabla B}$ & $+0.120$ [$-0.032$, $+0.236$] ($-0.345$--$+0.480$) & $+0.036$ [$-0.069$, $+0.224$] ($-0.306$--$+0.419$) \\
    $|H| a$ & $-0.186$ [$-0.279$, $-0.108$] ($-0.540$--$+0.128$) & $-0.236$ [$-0.332$, $-0.124$] ($-0.542$--$+0.124$) \\
    $\tau_{\rm axis} a$ & $-0.132$ [$-0.190$, $-0.076$] ($-0.357$--$+0.076$) & $-0.123$ [$-0.157$, $-0.074$] ($-0.337$--$+0.073$) \\
    $H a$ & $+0.215$ [$+0.124$, $+0.262$] ($-0.066$--$+0.481$) & $+0.220$ [$+0.116$, $+0.269$] ($-0.079$--$+0.481$) \\
    $\iota_{\rm edge}/n_{fp}$ & $+0.072$ [$-0.077$, $+0.137$] ($-0.273$--$+0.265$) & $+0.099$ [$-0.037$, $+0.167$] ($-0.240$--$+0.258$) \\
    \bottomrule
  \end{tabular}
\end{table}

\begin{table}[htbp]
  \centering
  \caption{Correlation between each predictor and mean coil non-planarity, all $n_{fp}$. Feasible population, demeaned within $(\text{campaign},n_{fp})$; 92 campaign cells with $n\ge50$, $\sum n=22478$, 53--661 per cell. Each entry is the median [Q1, Q3] (min--max) over those cells. Surface fields are reduced by an area-weighted mean and made dimensionless with the minor radius; both signed and absolute reductions are given where the sign carries meaning. Rows are ordered by $|$median $\rho|$ pooled over all cells.}
  \label{tab:pred-nfpall}
  \setlength{\tabcolsep}{6pt}
  \renewcommand{\arraystretch}{1.15}
  \small
  \begin{tabular}{lcc}
    \toprule
    predictor & Spearman $\rho$ & Pearson $r$ \\
    \midrule
    $\mathcal{T}$ & $+0.729$ [$+0.642$, $+0.774$] ($+0.430$--$+0.850$) & $+0.717$ [$+0.650$, $+0.777$] ($+0.403$--$+0.845$) \\
    $\kappa_n a$ & $+0.650$ [$+0.582$, $+0.720$] ($+0.229$--$+0.809$) & $+0.636$ [$+0.568$, $+0.707$] ($+0.315$--$+0.833$) \\
    $|K| a^2$ & $+0.464$ [$+0.164$, $+0.567$] ($-0.312$--$+0.686$) & $+0.439$ [$+0.258$, $+0.563$] ($-0.163$--$+0.695$) \\
    mirror$_{\rm edge}$ & $+0.333$ [$+0.253$, $+0.400$] ($+0.090$--$+0.591$) & $+0.366$ [$+0.287$, $+0.438$] ($+0.109$--$+0.655$) \\
    $|\tau_{\rm axis}| a$ & $+0.230$ [$+0.149$, $+0.373$] ($-0.015$--$+0.602$) & $+0.235$ [$+0.139$, $+0.335$] ($-0.051$--$+0.545$) \\
    $q_i$ & $+0.229$ [$+0.090$, $+0.401$] ($-0.222$--$+0.597$) & $+0.265$ [$+0.164$, $+0.404$] ($+0.009$--$+0.588$) \\
    $\kappa_{\rm axis} a$ & $+0.227$ [$-0.007$, $+0.508$] ($-0.491$--$+0.667$) & $+0.322$ [$+0.018$, $+0.453$] ($-0.458$--$+0.642$) \\
    $|\kappa_n| a$ & $+0.200$ [$-0.069$, $+0.361$] ($-0.510$--$+0.490$) & $+0.272$ [$+0.014$, $+0.367$] ($-0.435$--$+0.479$) \\
    vacuum well & $+0.177$ [$+0.086$, $+0.237$] ($-0.246$--$+0.557$) & $+0.164$ [$+0.043$, $+0.235$] ($-0.232$--$+0.481$) \\
    mirror$_{\rm axis}$ & $+0.169$ [$+0.090$, $+0.351$] ($-0.039$--$+0.583$) & $+0.196$ [$+0.136$, $+0.411$] ($-0.034$--$+0.615$) \\
    $\iota_{\rm axis}/n_{fp}$ & $-0.130$ [$-0.217$, $+0.150$] ($-0.308$--$+0.385$) & $-0.101$ [$-0.223$, $+0.114$] ($-0.335$--$+0.406$) \\
    $L_{\nabla B}$ & $-0.099$ [$-0.147$, $-0.003$] ($-0.367$--$+0.480$) & $-0.087$ [$-0.141$, $-0.009$] ($-0.367$--$+0.419$) \\
    $|H| a$ & $-0.096$ [$-0.178$, $+0.001$] ($-0.540$--$+0.486$) & $-0.080$ [$-0.140$, $+0.012$] ($-0.542$--$+0.488$) \\
    $\tau_{\rm axis} a$ & $-0.047$ [$-0.119$, $-0.005$] ($-0.357$--$+0.199$) & $-0.072$ [$-0.131$, $-0.032$] ($-0.393$--$+0.138$) \\
    $H a$ & $+0.031$ [$-0.042$, $+0.152$] ($-0.527$--$+0.481$) & $+0.016$ [$-0.072$, $+0.144$] ($-0.542$--$+0.481$) \\
    $\iota_{\rm edge}/n_{fp}$ & $-0.020$ [$-0.098$, $+0.070$] ($-0.273$--$+0.265$) & $-0.059$ [$-0.128$, $+0.048$] ($-0.274$--$+0.258$) \\
    \bottomrule
  \end{tabular}
\end{table}

\section{Correlation of \texorpdfstring{$\mathcal{T}$}{T} with the candidate quantities}
\label{app:taug-pairs}

Tables~\ref{tab:taug-axis}--\ref{tab:taug-bfield} give the coefficients behind
Figs.~\ref{fig:tg-axis}--\ref{fig:tg-bfield}, stratified by field-period count. These are
single correlations rather than summaries over campaign cells: both quantities in each pair
are stage-one, so the population is one row per configuration and there is nothing to take a
median over. Field and axis quantities are reduced on $|\cdot|$, by an area-weighted mean or
an area-weighted $95$th percentile; the percentile replaces the raw maximum, which on the
storage grid is a single point and is resolution-dependent. The $n_{fp}=5$ column rests on
$25$ configurations and should be read as indicative only.

\begin{table}[htbp]
  \centering
  \caption{Correlation of $\mathcal{T}$ with the magnetic axis. Each cell is the Spearman coefficient $\rho$ over the feasible population, one row per configuration (deduplicated across campaigns, since both quantities are stage-one). Surface and axis quantities are reduced on $|\cdot|$, by an area-weighted mean $\langle\cdot\rangle$ or an area-weighted $95$th percentile, and made dimensionless with the minor radius. The last column demeans within $n_{fp}$ before pooling. Configurations per stratum --- $n_{fp}=2$: 535, $n_{fp}=3$: 963, $n_{fp}=4$: 346, $n_{fp}=5$: 27.}
  \label{tab:taug-axis}
  \setlength{\tabcolsep}{6pt}
  \renewcommand{\arraystretch}{1.2}
  \small
  \begin{tabular}{lccccc}
    \toprule
     & $n_{fp}=2$ & $n_{fp}=3$ & $n_{fp}=4$ & $n_{fp}=5$ & all \\
    \midrule
    $\langle|\tau_{\rm axis}|\rangle a$ & $+0.23$ & $+0.09$ & $+0.75$ & $+0.58$ & $+0.30$ \\
    $|\tau_{\rm axis}|_{95}\,a$ & $+0.14$ & $+0.01$ & $+0.53$ & $+0.21$ & $+0.17$ \\
    $|\iota_{\rm axis}|/n_{fp}$ & $+0.48$ & $-0.19$ & $+0.58$ & $+0.25$ & $+0.19$ \\
    \bottomrule
  \end{tabular}
\end{table}

\begin{table}[htbp]
  \centering
  \caption{Correlation of $\mathcal{T}$ with global and physics scalars. Each cell is the Spearman coefficient $\rho$ over the feasible population, one row per configuration (deduplicated across campaigns, since both quantities are stage-one). Surface and axis quantities are reduced on $|\cdot|$, by an area-weighted mean $\langle\cdot\rangle$ or an area-weighted $95$th percentile, and made dimensionless with the minor radius. The last column demeans within $n_{fp}$ before pooling. Configurations per stratum --- $n_{fp}=2$: 535, $n_{fp}=3$: 963, $n_{fp}=4$: 346, $n_{fp}=5$: 27.}
  \label{tab:taug-global}
  \setlength{\tabcolsep}{6pt}
  \renewcommand{\arraystretch}{1.2}
  \small
  \begin{tabular}{lccccc}
    \toprule
     & $n_{fp}=2$ & $n_{fp}=3$ & $n_{fp}=4$ & $n_{fp}=5$ & all \\
    \midrule
    $L_{\nabla B}$ & $-0.23$ & $-0.09$ & $-0.45$ & $-0.45$ & $-0.20$ \\
    $|\iota_{\rm edge}|/n_{fp}$ & $+0.29$ & $+0.06$ & $+0.21$ & $+0.52$ & $+0.18$ \\
    vacuum well & $-0.25$ & $+0.12$ & $+0.02$ & $-0.40$ & $+0.01$ \\
    max elongation & $+0.03$ & $+0.09$ & $+0.05$ & $+0.21$ & $+0.06$ \\
    \bottomrule
  \end{tabular}
\end{table}

\begin{table}[htbp]
  \centering
  \caption{Correlation of $\mathcal{T}$ with surface differential geometry. Each cell is the Spearman coefficient $\rho$ over the feasible population, one row per configuration (deduplicated across campaigns, since both quantities are stage-one). Surface and axis quantities are reduced on $|\cdot|$, by an area-weighted mean $\langle\cdot\rangle$ or an area-weighted $95$th percentile, and made dimensionless with the minor radius. The last column demeans within $n_{fp}$ before pooling. Configurations per stratum --- $n_{fp}=2$: 535, $n_{fp}=3$: 963, $n_{fp}=4$: 346, $n_{fp}=5$: 27.}
  \label{tab:taug-surface}
  \setlength{\tabcolsep}{6pt}
  \renewcommand{\arraystretch}{1.2}
  \small
  \begin{tabular}{lccccc}
    \toprule
     & $n_{fp}=2$ & $n_{fp}=3$ & $n_{fp}=4$ & $n_{fp}=5$ & all \\
    \midrule
    $\langle|H|\rangle a$ & $+0.17$ & $-0.21$ & $+0.15$ & $+0.25$ & $-0.02$ \\
    $|H|_{95}\,a$ & $+0.16$ & $-0.06$ & $-0.52$ & $-0.05$ & $-0.14$ \\
    $\langle|K|\rangle a^2$ & $+0.43$ & $+0.51$ & $+0.54$ & $+0.31$ & $+0.52$ \\
    $|K|_{95}\,a^2$ & $+0.21$ & $+0.30$ & $+0.31$ & $+0.03$ & $+0.28$ \\
    \bottomrule
  \end{tabular}
\end{table}

\begin{table}[htbp]
  \centering
  \caption{Correlation of $\mathcal{T}$ with boundary $B$-field geometry. Each cell is the Spearman coefficient $\rho$ over the feasible population, one row per configuration (deduplicated across campaigns, since both quantities are stage-one). Surface and axis quantities are reduced on $|\cdot|$, by an area-weighted mean $\langle\cdot\rangle$ or an area-weighted $95$th percentile, and made dimensionless with the minor radius. The last column demeans within $n_{fp}$ before pooling. Configurations per stratum --- $n_{fp}=2$: 535, $n_{fp}=3$: 963, $n_{fp}=4$: 346, $n_{fp}=5$: 27.}
  \label{tab:taug-bfield}
  \setlength{\tabcolsep}{6pt}
  \renewcommand{\arraystretch}{1.2}
  \small
  \begin{tabular}{lccccc}
    \toprule
     & $n_{fp}=2$ & $n_{fp}=3$ & $n_{fp}=4$ & $n_{fp}=5$ & all \\
    \midrule
    $\langle|\kappa_n|\rangle a$ & $+0.46$ & $+0.51$ & $+0.20$ & $+0.28$ & $+0.40$ \\
    $|\kappa_n|_{95}\,a$ & $+0.47$ & $+0.56$ & $-0.10$ & $+0.21$ & $+0.38$ \\
    $\langle|\kappa_g|\rangle a$ & $+0.23$ & $+0.35$ & $+0.71$ & $+0.18$ & $+0.42$ \\
    $|\kappa_g|_{95}\,a$ & $+0.26$ & $+0.52$ & $+0.60$ & $+0.24$ & $+0.55$ \\
    \bottomrule
  \end{tabular}
\end{table}

\end{document}